\documentclass[twocolumn,english,aps,superscriptaddress,floatfix,longbibliography,nobibnotes,nofootinbib]{revtex4-2}
\usepackage{mathptmx}

\usepackage[T1]{fontenc}
\usepackage{color}
\usepackage{babel}
\usepackage{amsmath}
\usepackage{graphicx}
\usepackage{esint}
\usepackage[unicode=true,pdfusetitle,
 bookmarks=true,bookmarksnumbered=false,bookmarksopen=false,
 breaklinks=true,pdfborder={0 0 0},pdfborderstyle={},backref=false,colorlinks=true]
 {hyperref}
\hypersetup{
 linkcolor=blue,citecolor=blue,urlcolor=blue}

\makeatletter

\usepackage{tikz}
\usetikzlibrary{calc}

\usepackage{dsfont}
\usepackage{orcidlink}

\makeatother

\begin{document}
\title{Smeared phase transition in the dissipative random quantum Ashkin-Teller
model}
\author{Pedro S. Farinas\orcidlink{0000-0003-2208-7226}}
\affiliation{Instituto de F\'isica de S\~ao Carlos, Universidade de S\~ao Paulo,
S\~ao Carlos, SP, 13560-970, Brazil.}
\author{Rajesh Narayanan\orcidlink{0000-0002-4312-8915}}
\affiliation{Department of Physics, Indian Institute of Technology Madras, Chennai
600036, India.}
\author{Jos\'{e} A. Hoyos\orcidlink{0000-0003-2752-2194}}
\affiliation{Instituto de F\'isica de S\~ao Carlos, Universidade de S\~ao Paulo,
S\~ao Carlos, SP, 13560-970, Brazil.}
\begin{abstract}
We study the effects of dissipation in the phase diagram of the random
quantum Ashkin-Teller model by means of a generalization of the strong-disorder
renormalization group combined with adiabatic renormalization. This
model has three phases and three quantum phase transitions. We demonstrate
that the combined effect of Ohmic dissipation and quenched disorder
smears two out of the three quantum phase transitions. Our analytical
theory allows us to understand why one of the phase transitions remains
sharp. This is due to a cancellation of the dissipation effects on
the nontrivial nature of the intertwined order parameter of one of
the phases.\\
\\
Published in \href{https://doi.org/10.1103/mnwg-hd8m}{Phys. Rev. B {\bf 113}, 125128 (2026)}\hfill{}
DOI: \href{https://doi.org/10.1103/mnwg-hd8m}{10.1103/mnwg-hd8m}
\end{abstract}
\maketitle

\section{Introduction}

Quenched disorder underlies a manifold of profound effects near quantum
phase transitions: For instance, disorder-induced inhomogeneities
lead to exotic universality classes such as those typified by the
infinite randomness critical point~\citep{fisher_1,fisher_2}, and
the strong off-critical Griffiths singularities~\citep{griffiths_original,mccoy-prl69}.
Furthermore, in certain situations these inhomogeneities could lead
to the  smearing~\citep{smeared_hoyos_1,smeared_hoyos_2} of the
 sharp quantum phase transition. The moorings of the effects described
above is centered around the so-called rare regions (RRs): arbitrarily
large regions in space which, due to a statistically rare configuration
of the quenched-disordered couplings, is locally ordered and weakly
coupled to the bulk. These RRs, characterized by slow dynamics, dominate
the low-energy behavior of physical observables~\citep{glass_1,glass_2},
(for reviews, see Refs.~\citep{vojta2006rare,igloi-review,vojta2013phases,igloi-monthus-review2}). 

In practice it is important to realize that real materials are seldom
described as closed systems. Thus it is of paramount importance to
understand how the phases and phase transitions get affected when
the system is connected to a dissipative reservoir. Initially, it
was realized  that sufficiently strong dissipation can completely
damp the fluctuations of a RR~\citep{millis-morr-schmalian-prl01,vojta-prl03,vojta-schmalian-prb05,vojta-hoyos-prl14},
thus establishing that the combined effect of dissipation and disorder
is nonperturbative and could round the sharp phase transition by smearing.
To quantify the resulting phase and phase transition, interactions
among the RRs should be taken into account and it was shown that this
could be accomplished via a the  strong-disorder renormalization-group
(SDRG) method~\citep{numerical_ising1,numerical_ising2}. Later,
 it was established that for the case of the random transverse-field
Ising model (RTFIM), in any dimension, the critical point is entirely
destroyed by smearing when dissipation is Ohmic or Sub-Ohmic~\citep{smeared_hoyos_1,smeared_hoyos_2}.
Physically, the smearing phenomenon occurs because dissipation freezes
the dynamics of sufficiently large RRs, independent of the bulk, i.e.,
each rare region orders independently from the bulk, leading to local
phase transitions that can take place at different values of the control
parameter. Thus the paramagnetic Griffiths phase becomes an inhomogeneously
ordered ferromagnetic phase. Many other numerical studies (either
SDRG or Monte Carlo) have confirmed this conclusion~\citep{vojta2013phases,igloi-monthus-review2}.
Notably, experimental evidence of smeared phase transitions has been
observed in the compounds Sr$_{1-x}$Ca$_{x}$RuO$_{3}$~\citep{experimental_1}
and CePd$_{1-x}$Rh$_{x}$~\citep{experimental_2}. More recently,
smeared phase transitions have also been identified in a numerical
study of strange metals~\citep{strangemetals}, and have also potentially
been seen in low-dimensional superconductors~\citep{kaur-etal-njp24}. 

This work belongs to the genre of problems that we have discussed
in the previous paragraph: Namely, we study the dissipative random
quantum Ashkin-Teller (RQAT) model. Besides the usual paramagnetic
(PM) and ferromagnetic (FM) phases, this model features an additional
phase, called the product phase, which exhibits a composite (or intertwined)
order~\citep{1981_ashkinteller}. We inquire whether the transitions
between this product phase and the other phases are smeared by the
combined effects of dissipation and disorder. Interestingly, we show
that Ohmic dissipation smears only the transition between the product
and FM phases, while the transition between product and PM phases
remains sharp. This conclusion is robust and persists across different
forms of system-bath couplings. Surprisingly, even though dissipation
is coupled to the original order parameter fields, it does not couple
generically to the intertwined order parameter field and, therefore,
does not damp the quantum fluctuations of the associated rare regions
that form near the transition line between the product and PM phases.

Finally, besides this fundamental motivation, it is worth mentioning
that the Ashkin-Teller model has applications in other contexts such
as to describe layers of atoms absorbed on surfaces~\citep{bak-etal-prl85},
current loops in high-$T_{c}$ superconductors~\citep{aji-varma-prl07},
and the elastic response of DNA molecules~\citep{chang-wang-zheng-ctp08}.
The understanding of dissipation in these contexts is another motivation
for our work.

In the remainder of this work, we review the effects of disorder on
the quantum Ashkin-Teller chain in Sec.~\ref{sec:Review}, introduce
the model studied in Sec.~\ref{sec:A-simple-dissipative}, and apply
the adiabatic renormalization-group and the strong-disorder renormalization-group
methods to our problem in Secs.~\ref{sec:Adiabatic-RG} and \ref{sec:SDRG},
respectively. The resulting phase diagram is then discussed in Sec.~\ref{sec:Phase-diagram}.
Finally, we give concluding remarks in Sec.~\ref{sec:Discussion-and-conclusions}.
Other possible models of dissipation are considered in the Appendix~\ref{sec:Another-way-to}
and we find that our conclusions remain qualitatively correct. 

\section{Review of the dissipationless model\label{sec:Review}}

\subsection{Original variables}

The Hamiltonian of the one-dimensional random quantum Ashkin-Teller
(RQAT) model is~\citep{1943_ashkinteller,1981_ashkinteller,2001_ashkinteller,2008_ashkinteller,2014_ashkinteller}
\begin{align}
H_{\text{S}}= & -\sum_{\nu=1}^{2}\sum_{i}\left(J_{i}S_{\nu,i}^{z}S_{\nu,i+1}^{z}+h_{i}S_{\nu,i}^{x}\right)\nonumber \\
 & -\sum_{i}\left(K_{i}S_{1,i}^{z}S_{1,i+1}^{z}S_{2,i}^{z}S_{2,i+1}^{z}+g_{i}S_{1,i}^{x}S_{2,i}^{x}\right),\label{eq:H-AT}
\end{align}
where $S_{\nu,i}^{z}$ and $S_{\nu,i}^{x}$ are Pauli matrices representing
local spin-1/2 moments. The first two terms on the right-hand-side
of Eq.~(\ref{eq:H-AT}) describe two independent random transverse-field
Ising (RTFI) chains whereas the remaining terms quantify the couplings
between them. The index $\nu=1,2$, referred to as the color index,
distinguishes the two chains and the index $i$ distinguishes different
lattice sites. The interactions $J_{i}$ and transverse fields $h_{i}$
are strictly positive, independent random variables. The coupling
strengths are characterized by the ratios $\epsilon_{h,i}=g_{i}/h_{i}$
and $\epsilon_{J,i}=K_{i}/J_{i}$. For simplicity, we assume that
the (bare) values $\epsilon_{h,i}=\epsilon_{J,i}=\epsilon_{I}>0$;
i.e., they are uniform and site independent. However, they flow under
the renormalization-group transformations and become random variables~\citep{2014_ashkinteller}.
The low-energy physical behavior of the model (\ref{eq:H-AT}) can
be divided into two regimes: the weak-coupling regime ($\epsilon_{I}<1$)
and the strong-coupling regime ($\epsilon_{I}>1$). In the low-energy
limit of the weak-coupling regime, the two Ising chains effectively
decouple~\citep{2001_ashkinteller,2008_ashkinteller} and the resulting
physics is that of two identical RTFI chains. Therefore, there is
a direct quantum phase transition between a ferromagnet (where the
quantum average $\left\langle S_{\nu,i}^{z}\right\rangle \neq0$)
and paramagnet (where $\left\langle S_{\nu,i}^{z}\right\rangle =0$).
The transition, which occurs when $\prod J_{i}=\prod h_{i}$, is of
infinite-randomness type and is surrounded by the Griffiths ferro-
and paramagnetic phases. In the strong-coupling regime $\epsilon_{I}>1$,
on the other hand, this transition is through an intervening product
phase exhibiting a composite order; i.e., $\left\langle S_{\nu,i}^{z}\right\rangle =0$,
but $\left\langle S_{1,i}^{z}S_{2,i}^{z}\right\rangle \neq0$. The
transition between the ferromagnet and the product phase is also of
infinite-randomness type surrounded by quantum Griffiths ferromagnetic
and product phases. By duality,\footnote{The Hamiltonian (\ref{eq:H-AT}) is invariant under the duality transformation:
$S_{\nu,i}^{z}S_{\nu,i+1}^{z}\rightarrow\tau_{\nu,i}^{x}$, $S_{\nu,i}^{x}\rightarrow\tau_{\nu,i}^{z}\tau_{\nu,i+1}^{z}$
(here, $\tau^{x}$ and $\tau^{z}$ are the dual Pauli operators),
$J_{i}\leftrightarrow h_{i}$, and $K_{i}\leftrightarrow g_{i}$.} the transition between the paramagnetic and product phases is in
the same universality class~\citep{2014_ashkinteller}. 

\subsection{Product variable}

This result is better illustrated by reformulating the problem in
terms of ``product'' variables $\boldsymbol{\sigma}_{i}$ (which are
also spin-1/2 operators represented by Pauli matrices) and one of
the ``original'' spin variable $\boldsymbol{\eta}_{i}$, defined as
\begin{equation}
\sigma_{i}^{z}=S_{1,i}^{z}S_{2,i}^{z},\;\eta_{i}^{z}=S_{1,i}^{z},\;\sigma_{i}^{z}\eta_{i}^{z}=S_{2,i}^{z},\label{eq:2}
\end{equation}
and
\begin{equation}
\eta_{i}^{x}=S_{1,i}^{x}S_{2,i}^{x},\;\sigma_{i}^{x}=S_{2,i}^{x},\;\sigma_{i}^{x}\eta_{i}^{x}=S_{1,i}^{x}.\label{eq:3}
\end{equation}
In these new variables, the Hamiltonian (\ref{eq:H-AT}) becomes
\begin{align}
H_{\text{S}}= & -\sum_{i}\left(K_{i}\sigma_{i}^{z}\sigma_{i+1}^{z}+h_{i}\sigma_{i}^{x}\right)-\sum_{i}\left(J_{i}\eta_{i}^{z}\eta_{i+1}^{z}+g_{i}\eta_{i}^{x}\right)\nonumber \\
 & -\sum_{i}\left(J_{i}\sigma_{i}^{z}\sigma_{i+1}^{z}\eta_{i}^{z}\eta_{i+1}^{z}+h_{i}\sigma_{i}^{x}\eta_{i}^{x}\right).\label{eq:HAT-product}
\end{align}
 From the first and second sums of the Hamiltonian (\ref{eq:HAT-product}),
it is clear that, when $\epsilon_{I}\gg1$ and sufficiently near to
the self-duality line $\left(\delta\equiv\left\langle \ln\left(h/J\right)\right\rangle \approx0\right)$,
the system is in a phase with long-range order in the $\boldsymbol{\sigma}$
variables ($\left\langle \sigma^{z}\right\rangle \neq0$) and disordered
in the $\boldsymbol{\eta}$ variables ($\left\langle \eta^{z}\right\rangle =0$).
This implies that long-range order develops in the product variables
$\left\langle S_{1}^{z}S_{2}^{z}\right\rangle \neq0$, while the spins
remain disordered in each individual color $\left\langle S_{1}^{z}\right\rangle =\left\langle S_{2}^{z}\right\rangle =0$~\citep{1943_ashkinteller,1981_ashkinteller}-
thus, the moniker ``product (PROD) phase''. A caveat is needed here.
The terminology $\boldsymbol{\sigma}_{i}$ and $\boldsymbol{\eta}_{i}$
as a product and original variables is due to the lack of a better
one. It is precise if one considers only the $z$ component of the
original spins {[}see Eq.~(\ref{eq:2}){]}. Regarding the $x$ component
{[}see Eq.~(\ref{eq:3}){]}, however, the terminology is the other
way round. Regarding the $y$ component, both could be regarded as
product variables. In sum, both $\boldsymbol{\sigma}_{i}$ and $\boldsymbol{\eta}_{i}$
are nontrivial compositions of the original spin variables. We stick
with this terminology because the $z$ component is the one defining
the order parameter field of the transition.

\subsection{Phase diagram: Dissipationless case\label{subsec:PD}}

\begin{figure}[t]
\centering{}\includegraphics[scale=0.8]{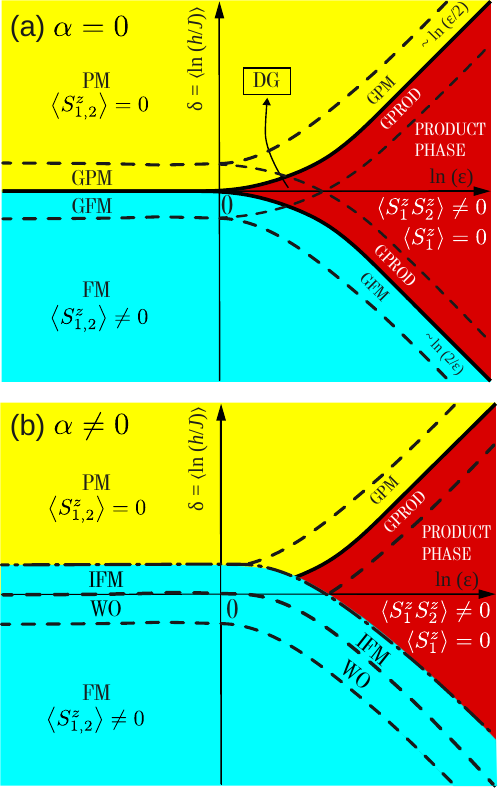}\caption{\label{fig:PD}(a) Schematic phase diagram of the RQAT chain without
dissipation ($\alpha=0$, discussed in Sec.~\ref{sec:Review}). 
GFM and GPM denote the Griffiths ferromagnetic and paramagnetic phases,
respectively. DG corresponds to the double Griffiths phase, characterized
by the presence of rare regions in both $\sigma$ and $\eta$ variables.
GPROD denotes the Griffiths product phase, where rare regions form
in the $\sigma$ variable near the PM transition and in the $\eta$
variable near the FM transition. (b) Schematic phase diagram of the
RQAT chain with dissipation ($\alpha\protect\neq0$, discussed in
Sec.~\ref{sec:Phase-diagram}). The transitions where rare regions
are formed in the original spin variable are smeared. In the strong-coupling
regime, the GFM phase is replaced by a weakly ordered (WO) phase and
the GPROD phase is replaced by an inhomogeneous ferromagnetic (IFM)
phase. In all cases, thick solid lines represent second-order quantum
phase transitions. The dot-dashed line in (b) represents the smeared
transition between the PROD and FM phases. The dashed lines indicate
crossovers, not transitions. Different colors represent different
phases.}
\end{figure}

In the presence of quenched disorder, it was found that all quantum
phase transitions belong to the same universality class (that of the
random transverse-field Ising chain) and are surrounded by the associated
quantum Griffiths phases~\citep{2014_ashkinteller} as illustrated
in the phase diagram Fig.~\hyperref[fig:PD]{\ref{fig:PD}(a)}.

It is important to distinguish between the two sorts of Griffiths
phases in the region $\epsilon_{I}>1$. Surrounding the PM-PROD transition,
the Griffiths singularities in both the Griffiths PM and Griffiths
PROD phases are due to rare regions locally in a PROD phase and weakly
interacting with the bulk. In RRs of this kind, the $z$ components
of the original spins fluctuate rapidly and independently of the nearby
spins. However, they fluctuate together with the other color spin
in the same site. On the other hand, surrounding the PROD-FM transition,
the low-energy behavior of both the Griffiths PROD and FM phases is
dominated by RRs locally in the FM phase and weakly coupled to the
bulk. Near the multicritical point, both RRs coexist in the region
$\epsilon_{I}>1$ and, therefore, give rise to the double Griffiths
(DG) phase~\citep{2014_ashkinteller}. It is worth noting that this
DG phase has been seen in a density-matrix renormalization-group study~\citep{chatelain-voliotis-epjb16}.

Evidently, these two phase transitions and the associated Griffiths
phases are related by duality. However, this will be broken when dissipation
is introduced as it couples to the original spin variables and not
to the product ones. As a result, only the FM-like RRs will be dramatically
affected by dissipation yielding to the smearing of the transition
as sketched in Fig.~\hyperref[fig:PD]{\ref{fig:PD}(b)} (see further
discussions in Sec.~\ref{sec:Phase-diagram}).

\section{Simple dissipative model\label{sec:A-simple-dissipative}}

We introduce dissipation by coupling each spin to its own set of identical
bosonic baths. The Hamiltonian thus becomes 
\begin{equation}
H=H_{\text{S}}+\sum_{\nu=1}^{2}H_{\text{B}}^{(\nu)}+\sum_{\nu=1}^{2}H_{\text{SB}}^{(\nu)}\label{eq:H-dissip}
\end{equation}
where $H_{\text{S}}$ is the RQAT Hamiltonian defined in Eq.~(\ref{eq:H-AT}).
The bath components are independent harmonic oscillators and, thus,
\begin{equation}
H_{\text{B}}^{(\nu)}=\sum_{k,i}\left(\frac{p_{\nu;k,i}^{2}}{2m_{k,i}}+\frac{m_{k,i}\omega_{k,i}^{2}x_{\nu;k,i}^{2}}{2}\right).
\end{equation}
Here, the oscillator frequencies $\omega_{k,i}$ and masses $m_{k,i}$
are assumed to be color independent. Finally, following the Caldeira-Leggett
model~\citep{caldeiraleggett,leggett_spinboson,smeared_hoyos_1,smeared_hoyos_2},
the system-bath coupling is 
\begin{equation}
H_{\text{SB}}^{(\nu)}=\sum_{i}S_{\nu,i}^{z}\sum_{k}\lambda_{k,i}x_{\nu;k,i},\label{eq:Hsb}
\end{equation}
 where the constant $\lambda_{k,i}$ controls the strength of the
coupling. For the moment, we consider color independent coupling constants.
Later, we discuss the effects of color dependence. 

The relevant properties of the baths are described by their spectral
densities $F_{i}(\omega)$, defined as
\begin{equation}
F_{i}(\omega)=\frac{\pi}{2}\sum_{k}\frac{\lambda_{k,i}^{2}}{m_{k,i}\omega_{k,i}}\delta(\omega-\omega_{k,i})=\frac{\pi}{2}\alpha_{i}\frac{\omega^{s}}{\omega_{c}^{s-1}}e^{-\omega/\omega_{c}}.\label{eq:spectral-F}
\end{equation}
Here, $\omega_{c}$ is the cutoff frequency, assumed to be identical
for all sites, and $\alpha_{i}$ is a dimensionless parameter quantifying
the dissipation strength. The exponent $s$ defines the dissipation
regime: $s=1$, $>1$, and $<1$ corresponding to the Ohmic, super-Ohmic,
and sub-Ohmic regimes, respectively.

Since we are interested in the $\epsilon>1$ regime, it is convenient
to reformulate the Hamiltonian using the product variables defined
in Eqs.~(\ref{eq:2}) and (\ref{eq:3}). The system-bath Hamiltonian
becomes
\begin{equation}
H_{\text{SB}}^{(1)}+H_{\text{SB}}^{(2)}=\sum_{i}\eta_{i}^{z}\sum_{k}\lambda_{k,i}x_{1;k,i}+\sum_{i}\sigma_{i}^{z}\eta_{i}^{z}\sum_{k}\lambda_{k,i}x_{2;k,i}.\label{eq:Hsb-new-variables}
\end{equation}
 Notice that the original variable $\boldsymbol{\eta}$ and the product
variable $\boldsymbol{\sigma}$ do not couple equally to the oscillator
baths. We will show that this has a profound consequence: the effects
of dissipation on the PM-PROD and PROD-FM phase transitions are fundamentally
different.

In what follows, we follow Refs.~\citep{smeared_hoyos_1,smeared_hoyos_2}
to generalize the adiabatic renormalization procedure~\citep{leggett_spinboson}
to the present case. Importantly, we assume weak dissipation ($\alpha_{i}\ll1$).
This means that, individually, each spin is weakly affected by dissipation.
Moreover, we also assume that the phases of the undamped system are
not unstable against weak dissipation, i.e., we assume that the nature
of the phases does not change by weak dissipation alone.

\section{Adiabatic renormalization\label{sec:Adiabatic-RG}}

This section applies the adiabatic renormalization procedure~\citep{leggett_spinboson}
to our model system. We remark that it is applicable to the regimes
of Ohmic and super-Ohmic dissipation $\left(s\ge1\right)$. Although
not applicable to the sub-Ohmic regime ($s<1$), insights can be gained.

The idea is to integrate out fast oscillators which can adiabatically
follow spins tunneling between the up and down $z$-states due to
$S^{x}$ operators in the system Hamiltonian (\ref{eq:HAT-product}).
Precisely, we integrate out the oscillators within the frequency range
$\omega_{l}<\omega_{k,i}<\omega_{c}$. The lower limit $\omega_{l}$
is defined as $\omega_{l}=pg_{i}=p\epsilon_{h,i}h_{i}$, where $p\gg1$
is a large real number the exact value of which is not important for
our purposes. Technically, $p\gg1$ ensures that we are integrating
out only oscillators with frequencies much higher than the tunneling
frequencies $g_{i}$ and $h_{i}$ and, thus, can instantaneously follow
to the spin dynamics and, hence, be adiabatically eliminated. The
net effect is a renormalization of the tunneling parameters $g_{i}$
and $h_{i}.$ To achieve this, we rewrite the system-bath parts of
the Hamiltonian by completing the square and separating out the high-frequency
components. The resulting Hamiltonian for the high-frequency oscillators
is
\begin{align}
H_{0}= & \sum_{\omega_{k,i}>\omega_{l}}\left(\frac{p_{1;k,i}^{2}}{2m_{k,i}}+\frac{m_{k,i}\omega_{k,i}^{2}\left(x_{1;k,i}+a_{0;k,i}\eta_{i}^{z}\right)^{2}}{2}\right)\nonumber \\
 & +\sum_{\omega_{k,i}>\omega_{l}}\left(\frac{p_{2;k,i}^{2}}{2m_{k,i}}+\frac{m_{k,i}\omega_{k,i}^{2}\left(x_{2;k,i}+a_{0;k,i}\sigma_{i}^{z}\eta_{i}^{z}\right)^{2}}{2}\right),
\end{align}
 where an irrelevant constant term has been omitted for simplicity
and $a_{0;k,i}=\lambda_{k,i}/m_{k,i}\omega_{k,i}^{2}$ is a constant.
Notice that $H_{0}$ is a collection of independent harmonic oscillators
with offset rest positions depending on $\eta_{i}^{z}$ and the other
on $\sigma_{i}^{z}\eta_{i}^{z}$ and, thus, its ground state is fourfold
degenerate: 
\begin{eqnarray}
\left|\Psi_{\pm\pm}\right\rangle  & = & \left|\pm,\pm\right\rangle \bigotimes_{\omega_{k,i}>\omega_{l}}\left|G_{k,i;\pm}^{(1)}\right\rangle \left|G_{k,i;\pm\pm}^{(2)}\right\rangle ,\\
\left|\Psi_{\pm\mp}\right\rangle  & = & \left|\pm,\mp\right\rangle \bigotimes_{\omega_{k,i}>\omega_{l}}\left|G_{k,i;\pm}^{(1)}\right\rangle \left|G_{k,i;\pm\mp}^{(2)}\right\rangle ,
\end{eqnarray}
where we are using the notation $\left|\eta_{i}^{z},\sigma_{i}^{z}\right\rangle $.
The oscillators' ground states have the form
\begin{equation}
\left|G_{k,i;\eta_{i}^{z}}^{(1)}\right\rangle =e^{i\eta_{i}^{z}p_{1;k,i}x_{0;k,i}}\left|0_{k,i}^{(1)}\right\rangle ,
\end{equation}
and
\begin{equation}
\left|G_{k,i;\eta_{i}^{z}\sigma_{i}^{z}}^{(2)}\right\rangle =e^{i\eta_{i}^{z}\sigma_{i}^{z}p_{2;k,i}x_{0;k,i}}\left|0_{k,i}^{(2)}\right\rangle ,
\end{equation}
with $\left|0_{k,i}^{(\nu)}\right\rangle $ denoting the ground state
of an oscillator with rest position at the origin. The tunneling term
in the Hamiltonian (\ref{eq:HAT-product}) is
\begin{equation}
H_{1}=-h_{i}\sigma_{i}^{x}\left(1+\eta_{i}^{x}\right)-g_{i}\eta_{i}^{x}.
\end{equation}
The other terms are all diagonal in the ground state and can simply
be incorporated to $H_{0}$ later on. Treating $H_{1}$ perturbatively
up to first order, we get the effective Hamiltonian
\begin{equation}
H_{\text{eff}}=-\tilde{h}_{i}\sigma_{i}^{x}\left(1+\eta_{i}^{x}\right)-\tilde{g}_{i}\eta_{i}^{x},\label{eq:Heff-adiabatic}
\end{equation}
with renormalized fields 
\begin{equation}
\tilde{h}_{i}=h_{i}e^{-\frac{2}{\pi}\intop_{p\epsilon_{h,i}h_{i}}^{\omega_{c}}d\omega\frac{F_{i}(\omega)}{\omega^{2}}}\approx h_{i}e^{-\alpha_{i}\intop_{p\epsilon_{h,i}h_{i}}^{\omega_{c}}d\omega\frac{\omega^{s-2}}{\omega_{c}^{s-1}}},\label{eq:h-AR}
\end{equation}
and
\begin{equation}
\tilde{g}_{i}=g_{i}e^{-\frac{4}{\pi}\intop_{pg_{i}}^{\omega_{c}}d\omega\frac{F_{i}(\omega)}{\omega^{2}}}\approx g_{i}e^{-2\alpha_{i}\intop_{pg_{i}}^{\omega_{c}}d\omega\frac{\omega^{s-2}}{\omega_{c}^{s-1}}}.\label{eq:g-AR}
\end{equation}
 Notice that the effective dissipation parameter for the product variables
(related to $g_{i}$) is twice that of the bare spin variables. This
may be expected since a spin flip in the product variable $\eta_{i}^{x}=S_{1,i}^{x}S_{2,i}^{x}$
involves changing the rest position of twice the number of oscillators
when compared to $\sigma_{i}^{x}=S_{1,i}^{x}$.

Because the effective Hamiltonian (\ref{eq:Heff-adiabatic}) keeps
its original form, this procedure can be iterated until convergence
of $g_{i}^{*}$ and $h_{i}^{*}$. As shown in Ref.~\citep{leggett_spinboson},
in the Ohmic regime ($s=1$), $h_{i}^{*}$ and $g_{i}^{*}$ converge
to finite values $h_{i}(p\epsilon_{h}h_{i}/\omega_{c})^{\alpha/\left(1-\alpha\right)}$
and $g_{i}(pg_{i}/\omega_{c})^{2\alpha/\left(1-2\alpha\right)}$ if
$\alpha_{i}<1$ and $<1/2$, respectively, otherwise, they converge
to vanishing values, and tunneling ceases. For super-Ohmic dissipation
($s>1$), $g_{i}^{*}$ and $h_{i}^{*}$ are always finite.

As we are interested in the case of weak dissipation, the bare value
of the dissipation parameter $\alpha_{i}\ll1$. Therefore, dissipation
has little effect individually as $h_{i}^{*}\approx h_{i}$ and $g_{i}^{*}\approx g_{i}$.
However, near a phase transition, spins at different sites become
correlated and, as shown in Ref.~\citep{smeared_hoyos_1}, the effective
dissipation parameter can change under RG. The central question then
becomes the following: under what conditions does the dissipation
parameter flow to larger values? As we show in the following, this
happens only when FM-like RRs appear, and in the FM phase - surprisingly,
not when PROD-like RRs are formed. 

\section{Strong-disorder renormalization group\label{sec:SDRG}}

We now present the SDRG decimation procedure relevant for the $\epsilon_{I}>1$
regime in the presence of dissipation. We follow the methods of the
dissipationless case in Ref.~\citep{2014_ashkinteller} and include
dissipation using the strategy of Ref.~\citep{smeared_hoyos_2} which
is the following. The renormalization-group decimations are of two
types: global and local, and are performed simultaneously when lowering
the cutoff energy scale of the system. The global decimation concerns
the integration of the fastest bath oscillators and is related to
the adiabatic renormalization (see Sec.~\ref{sec:Adiabatic-RG}).
The local decimation concerns the coarse grainning of a high-energy
localized mode in a disordered system and, thus, is identical to the
dissipationless case.

Let $\Omega=\mathrm{max}\left\{ K_{i},g_{i},J_{i},h_{i},\omega_{c}/p\right\} $
be the largest energy scale in the renormalized system. The decimation
procedure now unfolds in the following steps:

\subsection{Global decimation}

(i) In lowering the energy scale from $\Omega$ to $\Omega-d\Omega$,
we integrate out all the oscillators in the interval $\omega_{c}-pd\Omega<\omega_{k,i}<\omega_{c}=p\Omega$.
As a result, all tunneling frequencies renormalize to {[}see Eqs.~(\ref{eq:h-AR})
and (\ref{eq:g-AR}){]} 
\begin{align}
\tilde{h}_{i} & =h_{i}\left[1-\alpha_{i}\left(\frac{\Omega}{\Omega_{I}}\right)^{s-1}\frac{d\Omega}{\Omega}\right],\label{eq:h-tilde}
\end{align}
and
\begin{equation}
\tilde{g}_{i}=g_{i}\left[1-2\alpha_{i}\left(\frac{\Omega}{\Omega_{I}}\right)^{s-1}\frac{d\Omega}{\Omega}\right],\label{eq:g-tilde}
\end{equation}
 where $\Omega_{I}$ is the bare $\Omega$. Notice that this is the
global decimation procedure mentioned above. In contrast, the remaining
steps are local decimations and are identical to the dissipationless
case, except when it comes to renormalizing the dissipation parameter
$\alpha_{i}$. 

\subsection{Local decimations}

An interesting feature of the local decimations in the Ashkin-Teller
model, when compared to the Ising model, is that there are two coupled
dynamical variables in the problem: $\sigma$ and $\eta$ {[}see Eq.~(\ref{eq:HAT-product}){]}.
At zeroth order, they are not locally coupled and, therefore, the
decimation must take into account two local energy scales: one for
each variable. This is clear when one considers two decoupled Ising
models. A local decimation must take into account the largest energy
scale in each model.

\begin{figure}[b]
\begin{centering}
\includegraphics[clip,width=1\columnwidth]{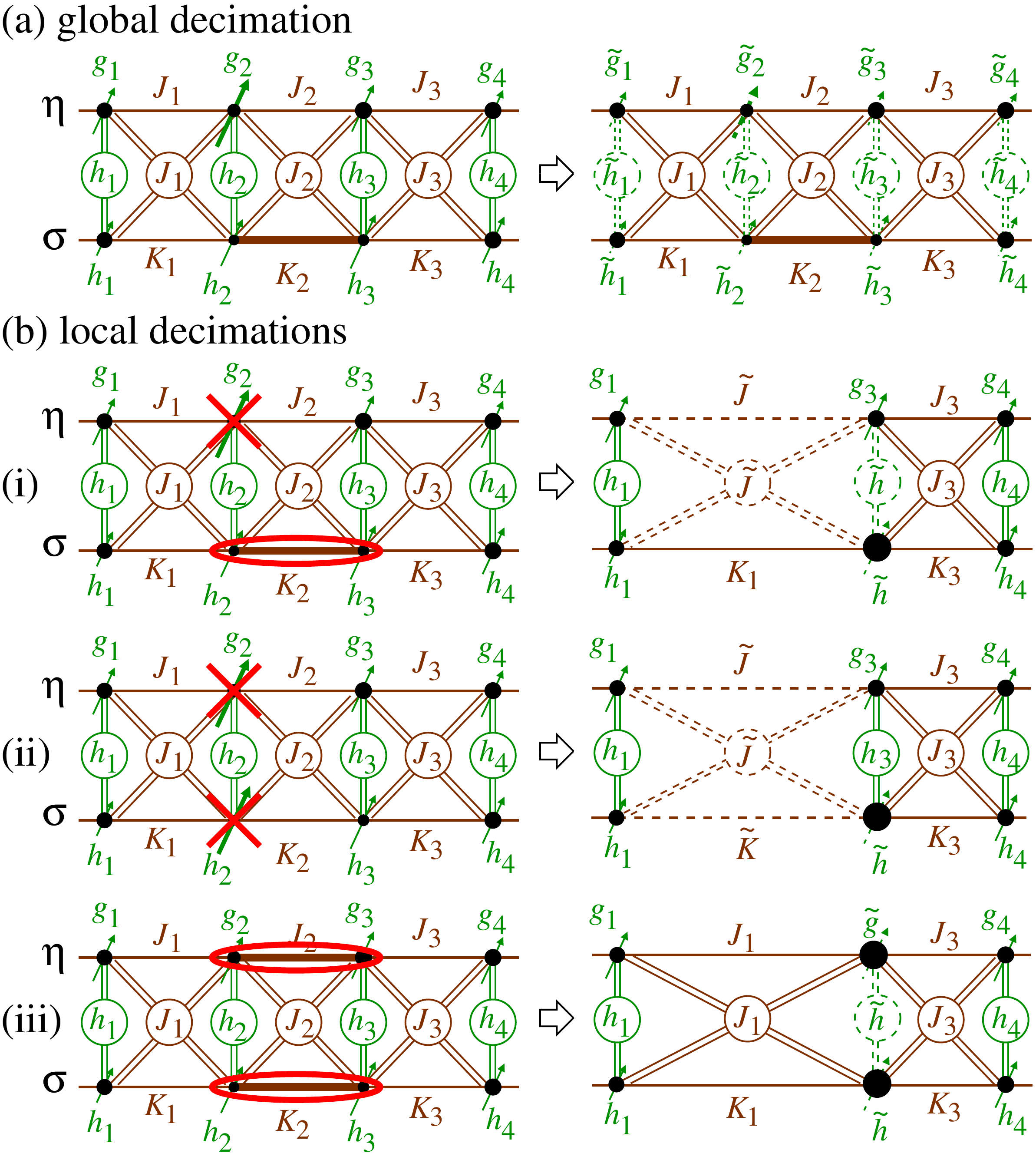}
\par\end{centering}
\caption{Decimation procedure: (a) Global decimation due to the coupling with
the dissipative bath {[}with $\tilde{h}_{i}$ and $\tilde{g}_{i}$
given by (\ref{eq:h-tilde}) and (\ref{eq:g-tilde}), respectively{]}.
(b) Local decimations due to a broad distribution of local energy
scales. When the two largest local energy scales in the system are
(i) $g_{2}$ and $K_{2}$ {[}with $\tilde{J}$ and $\tilde{h}$ given
by (\ref{eq:J-h-tilde}){]}, (ii) $g_{2}$ and $h_{2}$ {[}with $\tilde{K}$
and $\tilde{J}$ given by (\ref{eq:K-J-tilde}){]}, and (iii) $K_{2}$
and $J_{2}$ {[}with $\tilde{g}$ and $\tilde{h}$ given by (\ref{eq:g-h-tilde}){]}
In the main text, these correspond to the decimation cases (i), (ii),
(iii) and (iv), respectively. The couplings between the spin clusters
and the oscillator baths are also renormalized and are not illustrated
here.\label{fig:decimation}}

\end{figure}

(ii) Let the  largest energy scale in the system be $g_{i}$   and
the second largest in that local cluster be $K_{i}$ (or vice versa).
Thus the unperturbed Hamiltonian is $H_{0}=-K_{i}\sigma_{i}^{z}\sigma_{i+1}^{z}-g{}_{i}\eta_{i}^{x}$
and all other terms in the Hamiltonian (\ref{eq:HAT-product}) containing
$\boldsymbol{\sigma}_{i}$, $\boldsymbol{\sigma}_{i+1}$, and $\boldsymbol{\eta}_{i}$
are treated perturbatively up to second order. As a result, the new
effective Hamiltonian has the same form as the Hamiltonian (\ref{eq:HAT-product})
but with one fewer site and two fewer spin-1/2 degrees of freedom.
The degrees of freedom $\boldsymbol{\sigma}_{i}$ and $\boldsymbol{\sigma}_{i+1}$
are clustered together and replaced by an effective spin-1/2 degree
of freedom $\tilde{\boldsymbol{\sigma}}$ upon which the effective
field is $\tilde{h}$. In addition, $\boldsymbol{\eta}_{i}$ becomes
locked in the $x$ direction and does not contribute to the $z$ magnetization.
The effective coupling between the neighboring $\boldsymbol{\eta}$'s
is $\tilde{J}$. This is sketched in Fig.~\hyperref[fig:decimation]{\ref{fig:decimation}(a)},
and the renormalized couplings are 
\begin{equation}
\tilde{J}=\frac{2J_{i-1}J_{i}}{\tilde{g}_{i}}\mbox{ and }\tilde{h}=\frac{2h_{i}h_{i+1}}{K_{i}}.\label{eq:J-h-tilde}
\end{equation}
 Now we need to compute how the dissipation strength $\alpha$ changes
under this decimation. We recall that 
\begin{align}
H_{\text{SB}}= & -\eta_{i}^{z}\sum_{k}\lambda_{k,i}x_{1;k,i}-\sigma_{i}^{z}\eta_{i}^{z}\sum_{k}\lambda_{k,i}x_{2;k,i}\nonumber \\
 & -\sigma_{i+1}^{z}\eta_{i+1}^{z}\sum_{k}\lambda_{k,i+1}x_{2;k,i+1},\label{eq:HSB-decimation}
\end{align}
 and that the doubly degenerate ground state is $\left\{ \left|\rightarrow_{i};\uparrow_{i}\uparrow_{i+1}\right\rangle ,\left|\rightarrow_{i};\downarrow_{i}\downarrow_{i+1}\right\rangle \right\} $.
Then, projecting $H_{\text{SB}}$ onto these states, we find that
\begin{align}
\tilde{H}_{\text{SB}}= & -\tilde{\sigma}^{z}\eta_{i+1}^{z}\sum_{k}\lambda_{k,i+1}x_{2;k,i+1}.
\end{align}
This is simply because $\left\langle \eta_{i}^{z}\right\rangle =\left\langle \eta_{i}^{z}\sigma_{i}^{z}\right\rangle =0$
and $\left\langle \sigma_{i+1}^{z}\right\rangle =\tilde{\sigma}^{z}$.
Actually, this could be obtained in another fashion. Since $g_{i}$
is the largest energy scale, it means that the associated degree of
freedom is actually already (or nearly) decoupled from the associated
bath via successive adiabatic renormalizations (see Sec.~\ref{sec:Adiabatic-RG}).
Thus only the last term in (\ref{eq:HSB-decimation}) needs to be
projected onto the ground-state manifold. Importantly, there is no
change in the spectral function of the remaining baths and, therefore,
$\alpha_{i+1}$ remains unchanged.

(iii) Let the  largest energy scales be $g_{i}$ and the second largest
in that local cluster be $h_{i}$. In this case, all tunneling frequencies
have fully (or nearly) converged to their fixed-point values $g^{*}$
and $h^{*}$ in the adiabatic renormalization. The SDRG procedure
is, thus, the same as the dissipationless case. The unperturbed Hamiltonian
to be considered is $H_{0}=-h_{i}\sigma_{i}^{x}\left(1+\eta_{i}^{x}\right)-g_{i}\eta_{i}^{x}$.
Projecting the spins $\boldsymbol{\sigma}_{i}$ and $\boldsymbol{\eta}_{i}$
onto the ground state of $H_{0}$ means that both spins are going
to be decimated. The renormalized system is that illustrated in Fig.~\hyperref[fig:decimation]{\ref{fig:decimation}(b)}
and the renormalized quantities are
\begin{equation}
\tilde{K}=\frac{K_{i-1}K_{i}}{2h_{i}}\mbox{ and }\tilde{J}=\frac{J_{i-1}J_{i}}{g_{i}+h_{i}}.\label{eq:K-J-tilde}
\end{equation}
Since no spin clusters are formed in this step, the projection of
$H_{\text{SB}}$ into the ground state results in zero. Hence, $\alpha$
remains unchanged (except, evidently, for those ones decimated).

(iv) The final case to consider is when $K_{i}$ is the largest energy
scale in the system and $J_{i}$ is the second largest in that local
cluster. The unperturbed Hamiltonian is $H_{0}=-K_{i}\sigma_{i}^{z}\sigma_{i+1}^{z}-J_{i}\eta_{i}^{z}\eta_{i+1}^{z}-J_{i}\sigma_{i}^{z}\sigma_{i+1}^{z}\eta_{i}^{z}\eta_{i+1}^{z}$.
Projecting $\boldsymbol{\sigma}_{i}$, $\boldsymbol{\sigma}_{i+1}$,
$\boldsymbol{\eta}_{i}$, and $\boldsymbol{\eta}_{i+1}$ onto the
ground state means that those spins are always aligned and, therefore,
can be represented by two effective degrees of freedom $\tilde{\boldsymbol{\sigma}}$
and $\tilde{\boldsymbol{\eta}}$. Fusing them into clusters is illustrated
in Fig.~\hyperref[fig:decimation]{\ref{fig:decimation}(c)} where
the renormalized quantities are 
\begin{equation}
\tilde{g}=\frac{g_{i}g_{i+1}}{2J_{i}}\mbox{ and }\tilde{h}=\frac{h_{i}h_{i+1}}{K_{i}+J_{i}}.\label{eq:g-h-tilde}
\end{equation}
For this step, we need to take a closer look at what happens to the
oscillator baths. The system-bath Hamiltonian is 
\begin{align}
H_{\text{SB}}= & -\eta_{i}^{z}\sum_{k}\lambda_{k,i}x_{1;k,i}-\sigma_{i}^{z}\eta_{i}^{z}\sum_{k}\lambda_{k,i}x_{2;k,i}\nonumber \\
 & -\eta_{i+1}^{z}\sum_{k}\lambda_{k,i+1}x_{1;k,i+1}-\sigma_{i+1}^{z}\eta_{i+1}^{z}\sum_{k}\lambda_{k,i+1}x_{2;k,i+1},
\end{align}
 and needs to be projected onto the fourfold-degenerate ground state
of $H_{0}$, which is $\left\{ \left|\eta;\sigma\right\rangle \right\} $
(where $\eta_{j}^{z}\left|\eta;\sigma\right\rangle =\eta\left|\eta;\sigma\right\rangle $,
$\sigma_{j}^{z}\left|\eta;\sigma\right\rangle =\sigma\left|\eta;\sigma\right\rangle $,
with $j=i,i+1$ and $\eta,\sigma=\pm1$ representing the effective
spins). Since $H_{\text{SB}}$ is diagonal in all four states, the
projection can be easily performed, yielding the following effective
Hamiltonian
\begin{align}
\tilde{H}_{\text{SB}}= & -\tilde{\eta}^{z}\sum_{k,j}\lambda_{k,j}x_{1;k,j}-\tilde{\sigma}^{z}\tilde{\eta}^{z}\sum_{k,j}\lambda_{k,j}x_{2;k,j},
\end{align}
 where the sum in the index $j$ runs from $i$ to $i+1$. Consequently,
the spectral density of the local renormalized bath is 
\begin{equation}
\tilde{F}(\omega)=\frac{\pi}{2}\sum_{k,j}\frac{\lambda_{k,j}^{2}}{m_{k,j}\omega_{k,j}}\delta(\omega-\omega_{k,j})=\frac{\pi}{2}\tilde{\alpha}\frac{\omega^{s}}{\omega_{I}^{s-1}}e^{-\omega/\omega_{c}},\label{eq:F-tilde}
\end{equation}
 where we can identify the renormalized dissipation strength as 
\begin{equation}
\tilde{\alpha}=\alpha_{i}+\alpha_{i+1}.\label{eq:alfa-tilde}
\end{equation}
 It is thus interesting to notice that $\tilde{\alpha},$ increases
under the RG flow only when FM-like clusterings take place, i.e.,
only when $J_{i}$ is decimated. Moreover, from Eq.~(\ref{eq:alfa-tilde}),
it becomes clear that the effective dissipation strength increases
as the volume of the FM cluster, just like in the case of a direct
transition between the PM and FM phases studied in Refs.~\citep{smeared_hoyos_1,smeared_hoyos_2}.
In addition, it is possible to relate the dissipation strength with
the effective magnetic moment as both increase with the cluster volume.
It is a convenient relation when analyzing the SDRG flow equations
for the distributions of couplings, fields and dissipation parameters~\citep{smeared_hoyos_2}. 

\section{Phase diagram: dissipative case\label{sec:Phase-diagram}}

We are now able to describe the phase diagram of our model Hamiltonian
(\ref{eq:H-dissip}) sketched in Fig.~\ref{fig:PD}. A phase diagram
is obtained after analyzing the corresponding strong-disorder renormalization-group
flow of the Hamiltonian parameters. A careful reader may note that
the decimation cases (i)\textendash (iv) described in Sec.~\ref{sec:SDRG}
do not cover all possible decimations. However, they do cover all
decimations necessary to analyze the PM-PROD and PROD-FM transitions
as shown in Ref.~\citep{2014_ashkinteller}. The missing decimations
are those where $J_{i}$ or $h_{i}$ is the largest energy scale.
They are important in analyzing the $\epsilon<1$ case where the two
chains decouple and $\epsilon$ renormalizes to zero~\citep{2008_ashkinteller}.
This situation, thus, was analyzed elsewhere with and without dissipation~\citep{2008_ashkinteller,smeared_hoyos_1,smeared_hoyos_2} 

\subsection{Reviewing the dissipationless case}

To briefly review the undamped case (see further details in Sec.~\ref{subsec:PD}),
we start noticing that the PM-PROD transition happens from the competition
between the transverse field $h$ and the 4-spin coupling $K$ {[}see
the first sum in the Hamiltonian (\ref{eq:HAT-product}); the other
terms are just bystanders{]}. The relevant decimations are, thus,
those described in cases (ii) and (iii) in the previous section. Comparing
the corresponding renormalized values $\tilde{h}$ and $\tilde{K}$
in Eqs.~(\ref{eq:J-h-tilde}) and (\ref{eq:K-J-tilde}), the transition
line happens when the typical values are $2h_{\text{typ}}=K_{\text{typ}}$.
Thus the transition line in the undamped case is at $\delta\approx\epsilon_{I}/2$.
For the PROD-FM transition, the competition is between the 2-spin
couplings $J$ and $g$ {[}see the second sum in the Hamiltonian (\ref{eq:HAT-product}){]}.
The relevant decimations are thus the cases (ii) and (iv). Following
the same reasoning as before, the transition line is at $\delta\approx2/\epsilon_{I}.$

\subsection{Dissipative case: Sharp transition}

Let us now analyze the effects of weak Ohmic dissipation {[}$\alpha\ll1$
and $s=1$ in Eq.~(\ref{eq:spectral-F}){]} in the PM-PROD transition.
Since $\alpha\ll1$ and does not renormalize, the decimation of fast
oscillators has little effect on the renormalized values of $h$'s
{[}see Eq.~(\ref{eq:h-tilde}){]}. Therefore, Ohmic dissipation is
an irrelevant perturbation to the PM-PROD quantum phase transition.
Its only effect is a slight change of the critical line. This same
conclusion applies to the case of super-Ohmic dissipation ($s>1$)
which is qualitatively weaker when compared to Ohmic dissipation~\citep{leggett_spinboson}.
Very likely, the same conclusion may also apply to the case of sub-Ohmic
dissipation ($s<1$). 

\subsection{Dissipative case: Smeared transition}

The effects of dissipation are much more dramatic in the PROD-FM transition.
As pointed out in Secs.~\ref{sec:Review} and \ref{sec:SDRG}, FM-like
RRs are formed around that transition. This is because a local FM
coupling $J$ can be the second largest energy scale in the local
cluster {[}decimation case (iv) in Sec.~\ref{sec:SDRG}{]}. When
this happens, the effective dissipation strength $\tilde{\alpha}$
increases with the volume of the RR {[}see Eqs.~(\ref{eq:F-tilde})
and (\ref{eq:alfa-tilde}){]}. Thus, a sufficiently large RR (and
they can be arbitrarily large in a Griffiths phase~\citep{fisher_2})
has an effective dissipation parameter $\tilde{\alpha}>\alpha_{c}=1/2$.
As a consequence, these larger RRs will cease tunneling ($\tilde{g}\rightarrow0$)
and undergo their own FM transition when Ohmic dissipation is considered.
Thus, true local magnetic order develops. As this transition is not
a collective effect of the bulk, the transition is said to be smeared.
Any residual interaction between these larger RRs will align them
together and the total magnetization will not average out. The Griffiths
PROD phase of the undamped system now becomes an inhomogeneous FM
phase due to Ohmic dissipation. Upon lowering $\delta$, this inhomogeneous
FM phase crosses over to a weakly ordered phase (the former Griffiths
FM phase of the undamped system) and, finally, crosses over to the
conventional FM phase when further lowering $\delta.$ Having the
different names for the IFM and WO phases is to emphasize that in
the former, the magnetization appears only in some frozen and large
rare regions while in the latter, the magnetization happens in the
bulk. It is important to notice that this phenomenon is akin to the
smearing in the PM-FM transition in the random transverse-field Ising
chain studied in Refs.~\citep{smeared_hoyos_1,smeared_hoyos_2}.
Thus, both PM-FM ($\epsilon_{I}<1$) and PROD-FM ($\epsilon_{I}>1$)
transitions become smeared in the presence of Ohmic dissipation as
sketched in Fig.~\hyperref[fig:PD]{\ref{fig:PD}(b)}. We expect that
all these conclusions remain valid for sub-Ohmic dissipation as well.
For weak super-Ohmic dissipation, on the other hand, all transitions
remain sharp because a RR cannot stop tunneling due to the coupling
to the oscillator bath regardless the effective value of $\tilde{\alpha}$.

\subsection{Observables}

It is interesting to quantify some key observables in those transitions.
For the sharp transitions, the universality class is that of the random
transverse-field Ising model which is extensively studied. All critical
exponents are universal (i.e., independent of the details of the distributions
of the coupling constants and fields) and well known. We refer the
reader to Ref.~\citep{2014_ashkinteller} for the critical behavior
of many observables. For the smearing transitions, many observables
were quantified in Refs.~\citep{smeared_hoyos_1,smeared_hoyos_2}.
Here, we only quote the magnetization at the onset of the inhomogeneous
FM phase: $\overline{\left\langle \sigma_{i}^{z}\right\rangle }=\overline{\left\langle S_{1,i}^{z}\right\rangle }=\overline{\left\langle S_{2,i}^{z}\right\rangle }\sim w^{1/\alpha},$
where $\overline{\left\langle \cdot\right\rangle }$ denotes the disorder
and quantum averages. Here, $\alpha$ is the bare dissipation strength
and $w$ is the probability of finding a local coupling constant $J_{i}$
being greater than $\gamma_{i}=\max\left\{ h_{i}^{*},g_{i}^{*}\right\} $,
where $h^{*}=h\left(h/\Omega_{I}\right)^{\frac{\alpha}{1-\alpha}}\approx h$
and, likewise, $g^{*}\approx g$. Precisely, $w=\int_{0}^{\infty}dJP_{J}(J)\int_{0}^{J}d\gamma P_{\gamma}\left(\gamma\right)$.
Notice that $w$ plays the role of distance from transition.

\section{Discussions and conclusions \label{sec:Discussion-and-conclusions}}

In this work, we investigated the effects of dissipation in the phase
transitions of the quantum Ashkin-Teller chain in the presence of
quenched disorder. The dissipationless model exhibits three quantum
phase transitions {[}see Fig.~\hyperref[fig:PD]{\ref{fig:PD}(a)}{]}:
a FM-PM transition, a FM-PROD transition, and a PM-PROD transition,
all belonging to the universality class of the random transverse-field
Ising chain and displaying off-critical quantum Griffiths phases.
Introducing dissipation of the quantum fluctuations, surprisingly,
does not smear all of these phase transitions. Only the transitions
involving the FM phase are smeared {[}see Fig.~\hyperref[fig:PD]{\ref{fig:PD}(b)}{]}.

In the more familiar model of the random transverse-field Ising chain,
the low-energy physics is dominated by the so-called rare regions
in the Griffiths phases. These are regions which are locally FM and
weakly coupled to the bulk. Their dynamics are, thus, slow coherent
tunneling between the two lowest energy FM states. Smearing due to
dissipation occurs because that slow dynamics is completely halted
by Ohmic and sub-Ohmic dissipation. The coherent tunneling requires
a coherent changing in all oscillators' rest position. When that set
of oscillators is sufficiently slow, the rare region stops tunneling
and develops its own magnetic order regardless of the bulk. 

It turns out that the slow tunneling of the rare regions in the PM-PROD
phase does not require coherent tunneling of all their constituents.
This is because the magnetic order of the PROD phase is a composite
one: each color spin is disordered, $\left\langle S_{1}^{z}\right\rangle =\left\langle S_{2}^{z}\right\rangle =0$,
while the composite magnetic order finite, $\left\langle S_{1}^{z}S_{2}^{z}\right\rangle \neq0$.
Thus dissipation does not couple directly to the PROD order parameter
and, therefore, the transition remains sharp.

It is interesting to consider whether similar situations appear in
other contexts. Composite (quadrupolelike) nematic order appears in
diverse contexts such as in frustrated magnets (as, for instance,
in the context of order by disorder in the $J_{1}$-$J_{2}$ Heisenberg
model or in easy plane pyrochlores) and in iron pnictides. Giving
the results here uncovered that dissipation due to the coupling to
gapless modes (such as that of a metallic Fermi liquid) may not couple
to that composite magnetic order, and, consequently, the fluctuations
of the associated nematic rare regions may not be overdamped by dissipation. 

It is also interesting to test our result in other models. In the
Appendix~\ref{sec:Another-way-to} we consider two possible modifications
on the dissipation of the quantum fluctuations: one in which the couplings
to the oscillator bath are color dependent and another in which there
is only one oscillator bath for both color spins. In both cases, our
result remains: the smearing occurs only to the transitions to the
FM phase.

Finally, we argue that our results are not restricted to one spatial
dimension. Our SDRG method can be applied to any dimension and the
smearing (or lack thereof) will not depend on the underlying coordination
number.
\begin{acknowledgments}
This work was supported in part by the Brazilian agencies FAPESP and
CNPq. J.A.H. thanks IIT Madras for a visiting position under the IoE
program which facilitated the completion of this research work. R.N.
acknowledges funding from the Center for Quantum Information Theory
in Matter and Space-time, IIT Madras, and from the Department of Science
and Technology, Government of India, under Grant No. DST/ICPS/QuST/Theme-3/2019/Q69,
as well as support from the Mphasis F1 Foundation via the Centre for
Quantum Information, Communication, and Computing (CQuICC). 
\end{acknowledgments}

\appendix

\section{Other dissipative models\label{sec:Another-way-to}}

In this appendix, we consider two modifications of our model Hamiltonian
(\ref{eq:H-dissip}). The first modification (see App.~\ref{sec:Color-dependent-couplings})
is to introduce a color-dependent anisotropy in the couplings to the
different baths. The second modification is to consider that both
color spins at a given site couple to the same oscillator bath. 

We anticipate that our previous conclusions do not change: only the
transitions to the FM phase become smeared.

\subsection{Color-dependent couplings\label{sec:Color-dependent-couplings}}

Here, we consider the same model as in Sec.~\ref{sec:A-simple-dissipative}
but with color dependent coupling constants $\lambda_{k}^{(\nu)}$.
In this case, the spectral function will be color dependent $F_{i}^{(\nu)}$,
implying a color dependent dissipation strength $\alpha_{i}^{(\nu)}$.
The adiabatic renormalization procedure can be generalized, the main
difference being that the transverse field $h$ renormalizes differently
in each chain. Namely, $h$ unravels into $h^{(1)}$ and $h^{(2)}$
in the effective Hamiltonian
\begin{equation}
H_{\text{eff}}=-\tilde{h}_{i}^{(1)}\sigma^{x}-\tilde{h}_{i}^{(2)}\sigma^{x}\eta^{x}-\tilde{g}_{i}\eta^{x},
\end{equation}
with renormalized tunneling constants given by
\begin{equation}
\tilde{h}_{i}^{(1,2)}=h_{i}^{(1,2)}\exp\left(-\alpha_{i}^{(1,2)}\intop_{p'h_{i}^{(1,2)}}^{\omega_{c}}d\omega\frac{\omega^{s-2}}{\omega_{c}^{s-1}}\right)\label{eq:31}
\end{equation}
and
\begin{equation}
\tilde{g}_{i}=g_{i}\exp\left(-\left(\alpha_{i}^{(1)}+\alpha_{i}^{(2)}\right)\intop_{pg_{i}}^{\omega_{c}}d\omega\frac{\omega^{s-2}}{\omega_{c}^{s-1}}\right).\label{eq:32}
\end{equation}
Note that we recover the color independent result by setting $\alpha_{i}^{(1)}=\alpha_{i}^{(2)}$. 

It is necessary to check how the SDRG recursion relations from Sec.~\ref{sec:SDRG}
are modified by the unraveling of $h$. Eqs.~(\ref{eq:h-tilde})
and (\ref{eq:g-tilde}) in case (i) follows directly from relations
(\ref{eq:31}) and (\ref{eq:32}). In case (ii), $h^{(1)}$ and $h^{(2)}$
are renormalized in similar ways, namely Eq.~(\ref{eq:J-h-tilde})
is modified to $\tilde{h}^{(1,2)}=\left(h_{i}^{(1)}+h_{i}^{(2)}\right)h_{i+1}^{(1,2)}/K_{i}$,
and the recursion relation for $\tilde{J}$ remains unchanged. The
same applies to case (iv), with Eq.~(\ref{eq:g-h-tilde}) modified
to $\tilde{h}^{(1,2)}=h_{i}^{(1,2)}h_{i+1}^{(1,2)}/\left(K_{i}+J_{i}\right)$
and $\tilde{g}$ remains unchanged. Finally, in decimation case (iii),
Eq.~(\ref{eq:K-J-tilde}) modifies to 
\begin{equation}
\tilde{K}=\frac{K_{i-1}K_{i}}{h_{i}^{(1)}+h_{i}^{(2)}},\quad\tilde{J}^{(1,2)}=\frac{J_{i-1}J_{i}}{g_{i}+h_{i}^{(1,2)}}.
\end{equation}
 Notice, therefore, that $J$ acquires color-dependence, but not $K$.
Despite these modifications, the ground state of this step remains
the same as that of Sec.~\ref{sec:SDRG}. 

To check how dissipation will affect the phase transitions, we project
the coupling Hamiltonian $H_{\text{SB}}$ onto the ground states of
each decimation case. For cases (ii) and (iii), the coupling Hamiltonian
has the same form as the color-independent case (apart from the color
dependence of the $\lambda$'s). Therefore, there is no renormalization
of the dissipation strength $\alpha$. For case (iv), proceeding analogously
to Sec. \ref{sec:Phase-diagram}, we get 
\begin{equation}
H_{\text{eff}}=\tilde{\eta}^{z}\sum_{k}\tilde{\lambda}_{k}^{(1)}\tilde{x}_{1;k}+\tilde{\sigma}^{z}\tilde{\eta}^{z}\sum_{k}\tilde{\lambda}_{k}^{(2)}\tilde{x}_{2;k},
\end{equation}
where $\tilde{\lambda}_{k}^{(1,2)}=\lambda_{k,i}^{(1,2)}+\lambda_{k,i+1}^{(1,2)}$.
This result implies that the renormalization of the dissipation strength
in one color is independent of the other color, and they both renormalize
similarly to the color-independent case, with $\tilde{\alpha}^{(1,2)}=\alpha_{i}^{(1,2)}+\alpha_{i+1}^{(1,2)}$.
Hence, the color dependence of the coupling constant does not affect
our conclusions: only the transitions to the FM phase are smeared
due to Ohmic and sub-Ohmic dissipation.

\subsection{Single dissipation bath }

Here, we investigate whether our results on the model Hamiltonian
(\ref{eq:H-dissip}) also apply to another related dissipative model.
In (\ref{eq:H-dissip}), each spin is coupled to its own set of oscillator
baths. We now consider the case in which different color-spins are
coupled to the same set of oscillators. Namely, 
\begin{align}
H= & H_{\text{S}}+\sum_{i}\eta_{i}^{z}(1+\sigma_{i}^{z})\sum_{k}\lambda_{k,i}x_{k,i}\nonumber \\
 & +\sum_{k,i}\left(\frac{p_{k,i}^{2}}{2m_{k,i}}+\frac{m_{k,i}\omega_{k,i}^{2}x_{k,i}^{2}}{2}\right)+\triangle V.\label{eq:new-dissipH}
\end{align}
 The first term $H_{\text{S}}$ is our system model the quantum Ashkin-Teller
chain in Eq.~(\ref{eq:HAT-product}). The first sum quantifies the
coupling of the the two color spins to the same bath oscillator at
site $i$. {[}Recall that $\eta_{i}^{z}(1+\sigma_{i}^{z})=S_{1,i}^{z}+S_{3,i}^{z}$.{]}
The second sum is the total mechanical energy of the free oscillators
and, finally, 
\[
\triangle V=\frac{1}{2}\sum_{k,i}m_{k,i}\omega_{k,i}^{2}x_{0;k,i}^{2}\left(1+\sigma_{i}^{z}\right)^{2},
\]
 where $x_{0;k,i}=\lambda_{k,i}/m_{k,i}\omega_{k,i}^{2}$ is the counterterm~\citep{caldeiraleggett}.
It must be added if we wish that the coupling to the bath introduces
only dissipation, and not effective fields {[}which appears when completing
the square in the last line of Eq.~(\ref{eq:new-dissipH}){]}. To
proceed with the adiabatic renormalization, we set
\begin{align}
H_{0}= & \eta_{i}^{z}(1+\sigma_{i}^{z})\sum_{\omega_{k,i}>\omega_{l}}\lambda_{k,i}x_{k,i}+\sum_{\omega_{k,i}>\omega_{l}}\left(\frac{p_{k,i}^{2}}{2m_{k,i}}+\frac{m_{k,i}\omega_{k,i}^{2}x_{k,i}^{2}}{2}\right)\nonumber \\
 & +\sum_{\omega_{k,i}>\omega_{l}}\frac{m_{k,i}\omega_{k,i}^{2}x_{0;k,i}^{2}\left(1+\sigma_{i}^{z}\right)^{2}}{2}\nonumber \\
= & \sum_{\omega_{k,i}>\omega_{l}}\left(\frac{p_{k,i}^{2}}{2m_{k,i}}+\frac{m_{k,i}\omega_{k,i}^{2}\left(x_{k,i}+x_{0k,i}\eta_{i}^{z}(1+\sigma_{i}^{z})\right)^{2}}{2}\right),
\end{align}
 and treat the tunneling terms of the $i$th site in first order of
perturbation theory. The definition of $\omega_{l}$ is the same as
in Sec.~\ref{sec:Adiabatic-RG}. The fourfold degenerate ground state
of $H_{0}$ is given by
\begin{align}
\left|\Psi_{++}\right\rangle = & \left|+,+\right\rangle \bigotimes_{\omega_{k,i}>\omega_{l}}\left|G_{k,i;++}\right\rangle ,\nonumber \\
\left|\Psi_{-+}\right\rangle = & \left|-,+\right\rangle \bigotimes_{\omega_{k,i}>\omega_{l}}\left|G_{k,i;-+}\right\rangle ,\\
\left|\Psi_{\pm-}\right\rangle = & \left|\pm,-\right\rangle \bigotimes_{\omega_{k,i}>\omega_{l}}\left|0_{k,i}\right\rangle ,\nonumber 
\end{align}
 where we are using the same notation as Sec.~\ref{sec:Adiabatic-RG}
(first and second indices to $\eta$ and $\sigma$ variables, respectively).
The oscillators' ground states are 
\begin{equation}
\left|G_{k,i;\eta_{i}^{z}\sigma_{i}^{z}}\right\rangle =e^{2i\eta_{i}^{z}\sigma_{i}^{z}p_{k,i}x_{0;k,i}}\left|0_{k,i}\right\rangle ,
\end{equation}
 and $\left|0_{k,i}\right\rangle $ is the ground state of the oscillator
with the rest position at the origin. Projecting the perturbation
\begin{equation}
H_{1}=-h_{i}\sigma_{i}^{x}(1+\eta_{i}^{x})-g_{i}\eta_{i}^{x}\label{eq:40}
\end{equation}
 onto the ground state manyfold (the corresponding matrix element
being $H_{\text{eff}\eta^{z}\sigma^{z},\eta'^{z}\sigma'^{z}}=\left\langle \Psi_{\eta^{z}\sigma^{z}}\right|H_{1}\left|\Psi_{\eta'^{z}\sigma'^{z}}\right\rangle $),
we find the effective Hamiltonian 
\begin{align}
H_{\text{eff}}= & -\tilde{h_{i}}\sigma_{i}^{x}\left(1+\eta_{i}^{x}\right)-g_{i}\eta_{i}^{x}\left(\frac{1-\sigma_{i}^{z}}{2}\right)\nonumber \\
 & -\tilde{g_{i}}\eta_{i}^{x}\left(\frac{1+\sigma_{i}^{z}}{2}\right),\label{eq:42}
\end{align}
 with the renormalized tunneling frequencies 
\begin{equation}
\tilde{h}_{i}=h_{i}\exp\left(-\alpha_{i}\intop_{p'h_{i}}^{\omega_{c}}d\omega\frac{\omega^{s-2}}{\omega_{c}^{s-1}}\right),\label{eq:43}
\end{equation}
 and
\begin{equation}
\tilde{g}_{i}=g_{i}\exp\left(-4\alpha_{i}\intop_{pg_{i}}^{\omega_{c}}d\omega\frac{\omega^{s-2}}{\omega_{c}^{s-1}}\right).\label{eq:44}
\end{equation}
From Eq.~(\ref{eq:42}), it follows that the tunneling frequency
$g_{i}$ splits into two parts: one that decreases under adiabatic
renormalization and another that remains unaffected. This occurs because
both states $\left|\Psi_{\pm-}\right\rangle $ are centered at the
same position, such that the tunneling between states $\left|\Psi_{+-}\right\rangle $
and $\left|\Psi_{--}\right\rangle $, driven by the term $-g_{i}\eta_{i}^{x}$
in $H_{1}$, does not change the equilibrium position of the oscillators.
As a result, the tunneling frequency $g_{i}$ remains unaffected in
this case. However, the $-g_{i}\eta_{i}^{x}$ term also promotes tunneling
between the states $\left|\Psi_{++}\right\rangle $ and $\left|\Psi_{-+}\right\rangle $,
which are centered at different positions. For this tunneling process,
the frequency $g_{i}$ undergoes a renormalization, as described in
Eq.~(\ref{eq:44}). 

To iterate the procedure, we need to generalize the perturbation (\ref{eq:40})
as
\begin{align}
H_{1}= & -h_{i}\sigma_{i}^{x}\left(1+\eta_{i}^{x}\right)-g_{i}^{(a)}\eta_{i}^{x}\left(\frac{1-\sigma_{i}^{z}}{2}\right)\nonumber \\
 & -g_{i}^{(b)}\eta_{i}^{x}\left(\frac{1+\sigma_{i}^{z}}{2}\right)
\end{align}
 and start with $g_{i}^{(a)}=g_{i}^{(b)}$. In this way, the effective
Hamiltonian is
\begin{align}
H_{\text{eff}}= & -\tilde{h_{i}}\sigma_{i}^{x}\left(1+\eta_{i}^{x}\right)-g_{i}^{(a)}\eta_{i}^{x}\left(\frac{1-\sigma_{i}^{z}}{2}\right)\nonumber \\
 & -\tilde{g_{i}}^{(b)}\eta_{i}^{x}\left(\frac{1+\sigma_{i}^{z}}{2}\right),
\end{align}
 with $\tilde{h}_{i}$ and $\tilde{g}_{i}^{(b)}$ given, respectively,
by Eqs.~(\ref{eq:43}) and (\ref{eq:44}).

For the SDRG, we define the largest energy scale as $\Omega=\mathrm{max}\left\{ K_{i},\left(g_{i}^{(a)}+g_{i}^{(b)}\right)/2,J_{i},h_{i},\omega_{c}/p\right\} $,
and proceed just as in Sec.~\ref{sec:SDRG}. Decimation case (i)
follows directly from relations (\ref{eq:43}) and (\ref{eq:44}). 

For the decimation case (ii), the unperturbed Hamiltonian is $H_{0}=-K_{i}\sigma_{i}^{z}\sigma_{i+1}^{z}-\frac{\left(g_{i}^{(a)}+g_{i}^{(b)}\right)}{2}\eta_{i}^{x}$.
The additional term $\left(g_{i}^{(a)}-g_{i}^{(b)}\right)\eta_{i}^{x}\sigma_{i}^{z}/2$
(initially vanishing) is diagonal in the ground state of $H_{0}$
and contributes only with an unimportant constant, and the recursion
relations (\ref{eq:J-h-tilde}) become 
\begin{equation}
\tilde{J}=\frac{4J_{i-1}J_{i}}{g_{i}^{(a)}+g_{i}^{(b)}},\quad\tilde{h}=\frac{2h_{i}h_{i+1}}{K_{i}}.
\end{equation}
For the decimation case (iii), the unperturbed Hamiltonian is $H_{0}=-h_{i}\sigma_{i}^{x}\left(1+\eta_{i}^{x}\right)-\frac{\left(g_{i}^{(a)}+g_{i}^{(b)}\right)}{2}\eta_{i}^{x}$
and, adding the new term to $H_{1}$, the recursions (\ref{eq:K-J-tilde})
become
\begin{equation}
\tilde{K}=\frac{K_{i-1}K_{i}}{2h_{i}},\quad\tilde{J}=\frac{2J_{i-1}J_{i}}{2h_{i}+g_{i}^{(a)}+g_{i}^{(b)}}.
\end{equation}
For the decimation step (iv), nothing changes. The recursion relations
(\ref{eq:g-h-tilde}) remain but with a trivial modification of $g^{(a,b)}$.

Now, we can analyze the impact of dissipation on the phase transitions.
Since the steps of the SDRG process remain similar to those outlined
in Sec.~\ref{sec:SDRG}, the projection of 
\begin{equation}
H_{\text{SB}}=\eta_{i}^{z}(1+\sigma_{i}^{z})\sum_{k}\lambda_{k,i}x_{k,i}+\eta_{i+1}^{z}(1+\sigma_{i+1}^{z})\sum_{k}\lambda_{k,i+1}x_{k,i+1}
\end{equation}
onto the ground states corresponding to decimation cases (ii) and
(iii) does not alter the dissipation strength $\alpha$. In contrast,
for the case (iv), the coupling Hamiltonian becomes
\begin{equation}
\tilde{H}_{\text{SB}}=\tilde{\eta}^{z}(1+\tilde{\sigma}^{z})\sum_{k}\left(\lambda_{k,i}+\lambda_{k,i+1}\right)\tilde{x}_{k},
\end{equation}
which leads to the same renormalization of $\lambda_{k,i}$ (and,
consequently, $\alpha_{i}$) just like in the situation analyzed in
Sec.~\ref{sec:SDRG}. Thus, the effect of dissipation on the phase
transitions mirrors the earlier conclusions: the FM-PROD transition
is smeared, while the PM-PROD transition remains sharp.

\bibliography{/home/pedro/Documents/ashkin_teller_article/AT_arquives_publication/references_rel}

\begin{thebibliography}{33}%
\makeatletter
\providecommand \@ifxundefined [1]{%
 \@ifx{#1\undefined}
}%
\providecommand \@ifnum [1]{%
 \ifnum #1\expandafter \@firstoftwo
 \else \expandafter \@secondoftwo
 \fi
}%
\providecommand \@ifx [1]{%
 \ifx #1\expandafter \@firstoftwo
 \else \expandafter \@secondoftwo
 \fi
}%
\providecommand \natexlab [1]{#1}%
\providecommand \enquote  [1]{``#1''}%
\providecommand \bibnamefont  [1]{#1}%
\providecommand \bibfnamefont [1]{#1}%
\providecommand \citenamefont [1]{#1}%
\providecommand \href@noop [0]{\@secondoftwo}%
\providecommand \href [0]{\begingroup \@sanitize@url \@href}%
\providecommand \@href[1]{\@@startlink{#1}\@@href}%
\providecommand \@@href[1]{\endgroup#1\@@endlink}%
\providecommand \@sanitize@url [0]{\catcode `\\12\catcode `\$12\catcode
  `\&12\catcode `\#12\catcode `\^12\catcode `\_12\catcode `\%12\relax}%
\providecommand \@@startlink[1]{}%
\providecommand \@@endlink[0]{}%
\providecommand \url  [0]{\begingroup\@sanitize@url \@url }%
\providecommand \@url [1]{\endgroup\@href {#1}{\urlprefix }}%
\providecommand \urlprefix  [0]{URL }%
\providecommand \Eprint [0]{\href }%
\providecommand \doibase [0]{https://doi.org/}%
\providecommand \selectlanguage [0]{\@gobble}%
\providecommand \bibinfo  [0]{\@secondoftwo}%
\providecommand \bibfield  [0]{\@secondoftwo}%
\providecommand \translation [1]{[#1]}%
\providecommand \BibitemOpen [0]{}%
\providecommand \bibitemStop [0]{}%
\providecommand \bibitemNoStop [0]{.\EOS\space}%
\providecommand \EOS [0]{\spacefactor3000\relax}%
\providecommand \BibitemShut  [1]{\csname bibitem#1\endcsname}%
\let\auto@bib@innerbib\@empty
\bibitem [{\citenamefont {Fisher}(1994)}]{fisher_1}%
  \BibitemOpen
  \bibfield  {author} {\bibinfo {author} {\bibfnamefont {D.~S.}\ \bibnamefont
  {Fisher}},\ }\bibfield  {title} {\bibinfo {title} {Random antiferromagnetic
  quantum spin chains},\ }\href {https://doi.org/10.1103/PhysRevB.50.3799}
  {\bibfield  {journal} {\bibinfo  {journal} {Phys. Rev. B}\ }\textbf {\bibinfo
  {volume} {50}},\ \bibinfo {pages} {3799} (\bibinfo {year}
  {1994})}\BibitemShut {NoStop}%
\bibitem [{\citenamefont {Fisher}(1995)}]{fisher_2}%
  \BibitemOpen
  \bibfield  {author} {\bibinfo {author} {\bibfnamefont {D.~S.}\ \bibnamefont
  {Fisher}},\ }\bibfield  {title} {\bibinfo {title} {Critical behavior of
  random transverse-field ising spin chains},\ }\href
  {https://doi.org/10.1103/PhysRevB.51.6411} {\bibfield  {journal} {\bibinfo
  {journal} {Phys. Rev. B}\ }\textbf {\bibinfo {volume} {51}},\ \bibinfo
  {pages} {6411} (\bibinfo {year} {1995})}\BibitemShut {NoStop}%
\bibitem [{\citenamefont {Griffiths}(1969)}]{griffiths_original}%
  \BibitemOpen
  \bibfield  {author} {\bibinfo {author} {\bibfnamefont {R.~B.}\ \bibnamefont
  {Griffiths}},\ }\bibfield  {title} {\bibinfo {title} {Nonanalytic behavior
  above the critical point in a random ising ferromagnet},\ }\href
  {https://doi.org/10.1103/PhysRevLett.23.17} {\bibfield  {journal} {\bibinfo
  {journal} {Phys. Rev. Lett.}\ }\textbf {\bibinfo {volume} {23}},\ \bibinfo
  {pages} {17} (\bibinfo {year} {1969})}\BibitemShut {NoStop}%
\bibitem [{\citenamefont {McCoy}(1969)}]{mccoy-prl69}%
  \BibitemOpen
  \bibfield  {author} {\bibinfo {author} {\bibfnamefont {B.~M.}\ \bibnamefont
  {McCoy}},\ }\bibfield  {title} {\bibinfo {title} {Incompleteness of the
  critical exponent description for ferromagnetic systems containing random
  impurities},\ }\href {https://doi.org/10.1103/PhysRevLett.23.383} {\bibfield
  {journal} {\bibinfo  {journal} {Phys. Rev. Lett.}\ }\textbf {\bibinfo
  {volume} {23}},\ \bibinfo {pages} {383} (\bibinfo {year} {1969})}\BibitemShut
  {NoStop}%
\bibitem [{\citenamefont {Hoyos}\ and\ \citenamefont
  {Vojta}(2008)}]{smeared_hoyos_1}%
  \BibitemOpen
  \bibfield  {author} {\bibinfo {author} {\bibfnamefont {J.~A.}\ \bibnamefont
  {Hoyos}}\ and\ \bibinfo {author} {\bibfnamefont {T.}~\bibnamefont {Vojta}},\
  }\bibfield  {title} {\bibinfo {title} {Theory of smeared quantum phase
  transitions},\ }\href {https://doi.org/10.1103/PhysRevLett.100.240601}
  {\bibfield  {journal} {\bibinfo  {journal} {Phys. Rev. Lett.}\ }\textbf
  {\bibinfo {volume} {100}},\ \bibinfo {pages} {240601} (\bibinfo {year}
  {2008})}\BibitemShut {NoStop}%
\bibitem [{\citenamefont {Hoyos}\ and\ \citenamefont
  {Vojta}(2012)}]{smeared_hoyos_2}%
  \BibitemOpen
  \bibfield  {author} {\bibinfo {author} {\bibfnamefont {J.~A.}\ \bibnamefont
  {Hoyos}}\ and\ \bibinfo {author} {\bibfnamefont {T.}~\bibnamefont {Vojta}},\
  }\bibfield  {title} {\bibinfo {title} {Dissipation effects in random
  transverse-field ising chains},\ }\href
  {https://doi.org/10.1103/PhysRevB.85.174403} {\bibfield  {journal} {\bibinfo
  {journal} {Phys. Rev. B}\ }\textbf {\bibinfo {volume} {85}},\ \bibinfo
  {pages} {174403} (\bibinfo {year} {2012})}\BibitemShut {NoStop}%
\bibitem [{\citenamefont {Thill}\ and\ \citenamefont {Huse}(1995)}]{glass_1}%
  \BibitemOpen
  \bibfield  {author} {\bibinfo {author} {\bibfnamefont {M.}~\bibnamefont
  {Thill}}\ and\ \bibinfo {author} {\bibfnamefont {D.}~\bibnamefont {Huse}},\
  }\bibfield  {title} {\bibinfo {title} {Equilibrium behaviour of quantum ising
  spin glass},\ }\href
  {https://doi.org/https://doi.org/10.1016/0378-4371(94)00247-Q} {\bibfield
  {journal} {\bibinfo  {journal} {Physica A: Statistical Mechanics and its
  Applications}\ }\textbf {\bibinfo {volume} {214}},\ \bibinfo {pages} {321}
  (\bibinfo {year} {1995})}\BibitemShut {NoStop}%
\bibitem [{\citenamefont {Guo}\ \emph {et~al.}(1996)\citenamefont {Guo},
  \citenamefont {Bhatt},\ and\ \citenamefont {Huse}}]{glass_2}%
  \BibitemOpen
  \bibfield  {author} {\bibinfo {author} {\bibfnamefont {M.}~\bibnamefont
  {Guo}}, \bibinfo {author} {\bibfnamefont {R.~N.}\ \bibnamefont {Bhatt}},\
  and\ \bibinfo {author} {\bibfnamefont {D.~A.}\ \bibnamefont {Huse}},\
  }\bibfield  {title} {\bibinfo {title} {Quantum griffiths singularities in the
  transverse-field ising spin glass},\ }\href
  {https://doi.org/10.1103/PhysRevB.54.3336} {\bibfield  {journal} {\bibinfo
  {journal} {Phys. Rev. B}\ }\textbf {\bibinfo {volume} {54}},\ \bibinfo
  {pages} {3336} (\bibinfo {year} {1996})}\BibitemShut {NoStop}%
\bibitem [{\citenamefont {Vojta}(2006)}]{vojta2006rare}%
  \BibitemOpen
  \bibfield  {author} {\bibinfo {author} {\bibfnamefont {T.}~\bibnamefont
  {Vojta}},\ }\bibfield  {title} {\bibinfo {title} {Rare region effects at
  classical, quantum and nonequilibrium phase transitions},\ }\href@noop {}
  {\bibfield  {journal} {\bibinfo  {journal} {Journal of Physics A:
  Mathematical and General}\ }\textbf {\bibinfo {volume} {39}},\ \bibinfo
  {pages} {R143} (\bibinfo {year} {2006})}\BibitemShut {NoStop}%
\bibitem [{\citenamefont {Igl\'oi}\ and\ \citenamefont
  {Monthus}(2005)}]{igloi-review}%
  \BibitemOpen
  \bibfield  {author} {\bibinfo {author} {\bibfnamefont {F.}~\bibnamefont
  {Igl\'oi}}\ and\ \bibinfo {author} {\bibfnamefont {C.}~\bibnamefont
  {Monthus}},\ }\bibfield  {title} {\bibinfo {title} {Strong disorder {RG}
  approach of random systems},\ }\href
  {https://doi.org/10.1016/j.physrep.2005.02.006} {\bibfield  {journal}
  {\bibinfo  {journal} {Phys. Rep.}\ }\textbf {\bibinfo {volume} {412}},\
  \bibinfo {pages} {277} (\bibinfo {year} {2005})}\BibitemShut {NoStop}%
\bibitem [{\citenamefont {Vojta}(2013)}]{vojta2013phases}%
  \BibitemOpen
  \bibfield  {author} {\bibinfo {author} {\bibfnamefont {T.}~\bibnamefont
  {Vojta}},\ }\bibfield  {title} {\bibinfo {title} {Phases and phase
  transitions in disordered quantum systems},\ }in\ \href@noop {} {\emph
  {\bibinfo {booktitle} {AIP Conference Proceedings}}},\ Vol.\ \bibinfo
  {volume} {1550}\ (\bibinfo {organization} {American Institute of Physics},\
  \bibinfo {year} {2013})\ pp.\ \bibinfo {pages} {188--247}\BibitemShut
  {NoStop}%
\bibitem [{\citenamefont {Igl{\'o}i}\ and\ \citenamefont
  {Monthus}(2018)}]{igloi-monthus-review2}%
  \BibitemOpen
  \bibfield  {author} {\bibinfo {author} {\bibfnamefont {F.}~\bibnamefont
  {Igl{\'o}i}}\ and\ \bibinfo {author} {\bibfnamefont {C.}~\bibnamefont
  {Monthus}},\ }\bibfield  {title} {\bibinfo {title} {Strong disorder {RG}
  approach -- a short review of recent developments},\ }\href
  {https://doi.org/10.1140/epjb/e2018-90434-8} {\bibfield  {journal} {\bibinfo
  {journal} {The European Physical Journal B}\ }\textbf {\bibinfo {volume}
  {91}},\ \bibinfo {pages} {290} (\bibinfo {year} {2018})}\BibitemShut
  {NoStop}%
\bibitem [{\citenamefont {Millis}\ \emph {et~al.}(2001)\citenamefont {Millis},
  \citenamefont {Morr},\ and\ \citenamefont
  {Schmalian}}]{millis-morr-schmalian-prl01}%
  \BibitemOpen
  \bibfield  {author} {\bibinfo {author} {\bibfnamefont {A.~J.}\ \bibnamefont
  {Millis}}, \bibinfo {author} {\bibfnamefont {D.~K.}\ \bibnamefont {Morr}},\
  and\ \bibinfo {author} {\bibfnamefont {J.}~\bibnamefont {Schmalian}},\
  }\bibfield  {title} {\bibinfo {title} {Local defect in metallic quantum
  critical systems},\ }\href {https://doi.org/10.1103/PhysRevLett.87.167202}
  {\bibfield  {journal} {\bibinfo  {journal} {Phys. Rev. Lett.}\ }\textbf
  {\bibinfo {volume} {87}},\ \bibinfo {pages} {167202} (\bibinfo {year}
  {2001})}\BibitemShut {NoStop}%
\bibitem [{\citenamefont {Vojta}(2003)}]{vojta-prl03}%
  \BibitemOpen
  \bibfield  {author} {\bibinfo {author} {\bibfnamefont {T.}~\bibnamefont
  {Vojta}},\ }\bibfield  {title} {\bibinfo {title} {Disorder-induced rounding
  of certain quantum phase transitions},\ }\href
  {https://doi.org/10.1103/PhysRevLett.90.107202} {\bibfield  {journal}
  {\bibinfo  {journal} {Phys. Rev. Lett.}\ }\textbf {\bibinfo {volume} {90}},\
  \bibinfo {pages} {107202} (\bibinfo {year} {2003})}\BibitemShut {NoStop}%
\bibitem [{\citenamefont {Vojta}\ and\ \citenamefont
  {Schmalian}(2005)}]{vojta-schmalian-prb05}%
  \BibitemOpen
  \bibfield  {author} {\bibinfo {author} {\bibfnamefont {T.}~\bibnamefont
  {Vojta}}\ and\ \bibinfo {author} {\bibfnamefont {J.}~\bibnamefont
  {Schmalian}},\ }\bibfield  {title} {\bibinfo {title} {Quantum griffiths
  effects in itinerant heisenberg magnets},\ }\href
  {https://doi.org/10.1103/PhysRevB.72.045438} {\bibfield  {journal} {\bibinfo
  {journal} {Phys. Rev. B}\ }\textbf {\bibinfo {volume} {72}},\ \bibinfo
  {pages} {045438} (\bibinfo {year} {2005})}\BibitemShut {NoStop}%
\bibitem [{\citenamefont {Vojta}\ and\ \citenamefont
  {Hoyos}(2014)}]{vojta-hoyos-prl14}%
  \BibitemOpen
  \bibfield  {author} {\bibinfo {author} {\bibfnamefont {T.}~\bibnamefont
  {Vojta}}\ and\ \bibinfo {author} {\bibfnamefont {J.~A.}\ \bibnamefont
  {Hoyos}},\ }\bibfield  {title} {\bibinfo {title} {Criticality and quenched
  disorder: Harris criterion versus rare regions},\ }\href
  {https://doi.org/10.1103/PhysRevLett.112.075702} {\bibfield  {journal}
  {\bibinfo  {journal} {Phys. Rev. Lett.}\ }\textbf {\bibinfo {volume} {112}},\
  \bibinfo {pages} {075702} (\bibinfo {year} {2014})}\BibitemShut {NoStop}%
\bibitem [{\citenamefont {Schehr}\ and\ \citenamefont
  {Rieger}(2006)}]{numerical_ising1}%
  \BibitemOpen
  \bibfield  {author} {\bibinfo {author} {\bibfnamefont {G.}~\bibnamefont
  {Schehr}}\ and\ \bibinfo {author} {\bibfnamefont {H.}~\bibnamefont
  {Rieger}},\ }\bibfield  {title} {\bibinfo {title} {Strong-disorder fixed
  point in the dissipative random transverse-field ising model},\ }\href
  {https://doi.org/10.1103/PhysRevLett.96.227201} {\bibfield  {journal}
  {\bibinfo  {journal} {Phys. Rev. Lett.}\ }\textbf {\bibinfo {volume} {96}},\
  \bibinfo {pages} {227201} (\bibinfo {year} {2006})}\BibitemShut {NoStop}%
\bibitem [{\citenamefont {Schehr}\ and\ \citenamefont
  {Rieger}(2008)}]{numerical_ising2}%
  \BibitemOpen
  \bibfield  {author} {\bibinfo {author} {\bibfnamefont {G.}~\bibnamefont
  {Schehr}}\ and\ \bibinfo {author} {\bibfnamefont {H.}~\bibnamefont
  {Rieger}},\ }\bibfield  {title} {\bibinfo {title} {Finite temperature
  behavior of strongly disordered quantum magnets coupled to a dissipative
  bath*},\ }\href {https://doi.org/10.1088/1742-5468/2008/04/P04012} {\bibfield
   {journal} {\bibinfo  {journal} {Journal of Statistical Mechanics: Theory and
  Experiment}\ }\textbf {\bibinfo {volume} {2008}},\ \bibinfo {pages} {P04012}
  (\bibinfo {year} {2008})}\BibitemShut {NoStop}%
\bibitem [{\citenamefont {Demk\'o}\ \emph {et~al.}(2012)\citenamefont
  {Demk\'o}, \citenamefont {Bord\'acs}, \citenamefont {Vojta}, \citenamefont
  {Nozadze}, \citenamefont {Hrahsheh}, \citenamefont {Svoboda}, \citenamefont
  {D\'ora}, \citenamefont {Yamada}, \citenamefont {Kawasaki}, \citenamefont
  {Tokura},\ and\ \citenamefont {K\'ezsm\'arki}}]{experimental_1}%
  \BibitemOpen
  \bibfield  {author} {\bibinfo {author} {\bibfnamefont {L.}~\bibnamefont
  {Demk\'o}}, \bibinfo {author} {\bibfnamefont {S.}~\bibnamefont {Bord\'acs}},
  \bibinfo {author} {\bibfnamefont {T.}~\bibnamefont {Vojta}}, \bibinfo
  {author} {\bibfnamefont {D.}~\bibnamefont {Nozadze}}, \bibinfo {author}
  {\bibfnamefont {F.}~\bibnamefont {Hrahsheh}}, \bibinfo {author}
  {\bibfnamefont {C.}~\bibnamefont {Svoboda}}, \bibinfo {author} {\bibfnamefont
  {B.}~\bibnamefont {D\'ora}}, \bibinfo {author} {\bibfnamefont
  {H.}~\bibnamefont {Yamada}}, \bibinfo {author} {\bibfnamefont
  {M.}~\bibnamefont {Kawasaki}}, \bibinfo {author} {\bibfnamefont
  {Y.}~\bibnamefont {Tokura}},\ and\ \bibinfo {author} {\bibfnamefont
  {I.}~\bibnamefont {K\'ezsm\'arki}},\ }\bibfield  {title} {\bibinfo {title}
  {Disorder promotes ferromagnetism: Rounding of the quantum phase transition
  in ${\mathrm{sr}}_{1\ensuremath{-}x}{\mathrm{ca}}_{x}{\mathrm{ruo}}_{3}$},\
  }\href {https://doi.org/10.1103/PhysRevLett.108.185701} {\bibfield  {journal}
  {\bibinfo  {journal} {Phys. Rev. Lett.}\ }\textbf {\bibinfo {volume} {108}},\
  \bibinfo {pages} {185701} (\bibinfo {year} {2012})}\BibitemShut {NoStop}%
\bibitem [{\citenamefont {Sereni}\ \emph {et~al.}(2007)\citenamefont {Sereni},
  \citenamefont {Westerkamp}, \citenamefont {K\"uchler}, \citenamefont
  {Caroca-Canales}, \citenamefont {Gegenwart},\ and\ \citenamefont
  {Geibel}}]{experimental_2}%
  \BibitemOpen
  \bibfield  {author} {\bibinfo {author} {\bibfnamefont {J.~G.}\ \bibnamefont
  {Sereni}}, \bibinfo {author} {\bibfnamefont {T.}~\bibnamefont {Westerkamp}},
  \bibinfo {author} {\bibfnamefont {R.}~\bibnamefont {K\"uchler}}, \bibinfo
  {author} {\bibfnamefont {N.}~\bibnamefont {Caroca-Canales}}, \bibinfo
  {author} {\bibfnamefont {P.}~\bibnamefont {Gegenwart}},\ and\ \bibinfo
  {author} {\bibfnamefont {C.}~\bibnamefont {Geibel}},\ }\bibfield  {title}
  {\bibinfo {title} {Ferromagnetic quantum criticality in the alloy
  $\mathrm{Ce}{\mathrm{pd}}_{1\ensuremath{-}x}{\mathrm{rh}}_{x}$},\ }\href
  {https://doi.org/10.1103/PhysRevB.75.024432} {\bibfield  {journal} {\bibinfo
  {journal} {Phys. Rev. B}\ }\textbf {\bibinfo {volume} {75}},\ \bibinfo
  {pages} {024432} (\bibinfo {year} {2007})}\BibitemShut {NoStop}%
\bibitem [{\citenamefont {Patel}\ \emph {et~al.}(2024)\citenamefont {Patel},
  \citenamefont {Lunts},\ and\ \citenamefont {Sachdev}}]{strangemetals}%
  \BibitemOpen
  \bibfield  {author} {\bibinfo {author} {\bibfnamefont {A.~A.}\ \bibnamefont
  {Patel}}, \bibinfo {author} {\bibfnamefont {P.}~\bibnamefont {Lunts}},\ and\
  \bibinfo {author} {\bibfnamefont {S.}~\bibnamefont {Sachdev}},\ }\bibfield
  {title} {\bibinfo {title} {Localization of overdamped bosonic modes and
  transport in strange metals},\ }\href
  {https://doi.org/10.1073/pnas.2402052121} {\bibfield  {journal} {\bibinfo
  {journal} {Proceedings of the National Academy of Sciences}\ }\textbf
  {\bibinfo {volume} {121}},\ \bibinfo {pages} {e2402052121} (\bibinfo {year}
  {2024})},\ \Eprint
  {https://arxiv.org/abs/https://www.pnas.org/doi/pdf/10.1073/pnas.2402052121}
  {https://www.pnas.org/doi/pdf/10.1073/pnas.2402052121} \BibitemShut {NoStop}%
\bibitem [{\citenamefont {Kaur}\ \emph {et~al.}(2024)\citenamefont {Kaur},
  \citenamefont {Kumar~Kundu}, \citenamefont {Kumar}, \citenamefont {Dogra},
  \citenamefont {Narayanan}, \citenamefont {Vojta},\ and\ \citenamefont
  {Bid}}]{kaur-etal-njp24}%
  \BibitemOpen
  \bibfield  {author} {\bibinfo {author} {\bibfnamefont {S.}~\bibnamefont
  {Kaur}}, \bibinfo {author} {\bibfnamefont {H.}~\bibnamefont {Kumar~Kundu}},
  \bibinfo {author} {\bibfnamefont {S.}~\bibnamefont {Kumar}}, \bibinfo
  {author} {\bibfnamefont {A.}~\bibnamefont {Dogra}}, \bibinfo {author}
  {\bibfnamefont {R.}~\bibnamefont {Narayanan}}, \bibinfo {author}
  {\bibfnamefont {T.}~\bibnamefont {Vojta}},\ and\ \bibinfo {author}
  {\bibfnamefont {A.}~\bibnamefont {Bid}},\ }\bibfield  {title} {\bibinfo
  {title} {Novel emergent phases in a two-dimensional superconductor},\ }\href
  {https://doi.org/10.1088/1367-2630/ad6800} {\bibfield  {journal} {\bibinfo
  {journal} {New Journal of Physics}\ }\textbf {\bibinfo {volume} {26}},\
  \bibinfo {pages} {083001} (\bibinfo {year} {2024})}\BibitemShut {NoStop}%
\bibitem [{\citenamefont {Kohmoto}\ \emph {et~al.}(1981)\citenamefont
  {Kohmoto}, \citenamefont {den Nijs},\ and\ \citenamefont
  {Kadanoff}}]{1981_ashkinteller}%
  \BibitemOpen
  \bibfield  {author} {\bibinfo {author} {\bibfnamefont {M.}~\bibnamefont
  {Kohmoto}}, \bibinfo {author} {\bibfnamefont {M.}~\bibnamefont {den Nijs}},\
  and\ \bibinfo {author} {\bibfnamefont {L.~P.}\ \bibnamefont {Kadanoff}},\
  }\bibfield  {title} {\bibinfo {title} {Hamiltonian studies of the $d=2$
  ashkin-teller model},\ }\href {https://doi.org/10.1103/PhysRevB.24.5229}
  {\bibfield  {journal} {\bibinfo  {journal} {Phys. Rev. B}\ }\textbf {\bibinfo
  {volume} {24}},\ \bibinfo {pages} {5229} (\bibinfo {year}
  {1981})}\BibitemShut {NoStop}%
\bibitem [{\citenamefont {Bak}\ \emph {et~al.}(1985)\citenamefont {Bak},
  \citenamefont {Kleban}, \citenamefont {Unertl}, \citenamefont {Ochab},
  \citenamefont {Akinci}, \citenamefont {Bartelt},\ and\ \citenamefont
  {Einstein}}]{bak-etal-prl85}%
  \BibitemOpen
  \bibfield  {author} {\bibinfo {author} {\bibfnamefont {P.}~\bibnamefont
  {Bak}}, \bibinfo {author} {\bibfnamefont {P.}~\bibnamefont {Kleban}},
  \bibinfo {author} {\bibfnamefont {W.~N.}\ \bibnamefont {Unertl}}, \bibinfo
  {author} {\bibfnamefont {J.}~\bibnamefont {Ochab}}, \bibinfo {author}
  {\bibfnamefont {G.}~\bibnamefont {Akinci}}, \bibinfo {author} {\bibfnamefont
  {N.~C.}\ \bibnamefont {Bartelt}},\ and\ \bibinfo {author} {\bibfnamefont
  {T.~L.}\ \bibnamefont {Einstein}},\ }\bibfield  {title} {\bibinfo {title}
  {Phase diagram of selenium adsorbed on the ni(100) surface: A physical
  realization of the ashkin-teller model},\ }\href
  {https://doi.org/10.1103/PhysRevLett.54.1539} {\bibfield  {journal} {\bibinfo
   {journal} {Phys. Rev. Lett.}\ }\textbf {\bibinfo {volume} {54}},\ \bibinfo
  {pages} {1539} (\bibinfo {year} {1985})}\BibitemShut {NoStop}%
\bibitem [{\citenamefont {Aji}\ and\ \citenamefont
  {Varma}(2007)}]{aji-varma-prl07}%
  \BibitemOpen
  \bibfield  {author} {\bibinfo {author} {\bibfnamefont {V.}~\bibnamefont
  {Aji}}\ and\ \bibinfo {author} {\bibfnamefont {C.~M.}\ \bibnamefont
  {Varma}},\ }\bibfield  {title} {\bibinfo {title} {Theory of the quantum
  critical fluctuations in cuprate superconductors},\ }\href
  {https://doi.org/10.1103/PhysRevLett.99.067003} {\bibfield  {journal}
  {\bibinfo  {journal} {Phys. Rev. Lett.}\ }\textbf {\bibinfo {volume} {99}},\
  \bibinfo {pages} {067003} (\bibinfo {year} {2007})}\BibitemShut {NoStop}%
\bibitem [{\citenamefont {Zhe}\ \emph {et~al.}(2008)\citenamefont {Zhe},
  \citenamefont {Ping},\ and\ \citenamefont
  {Ying-Hong}}]{chang-wang-zheng-ctp08}%
  \BibitemOpen
  \bibfield  {author} {\bibinfo {author} {\bibfnamefont {C.}~\bibnamefont
  {Zhe}}, \bibinfo {author} {\bibfnamefont {W.}~\bibnamefont {Ping}},\ and\
  \bibinfo {author} {\bibfnamefont {Z.}~\bibnamefont {Ying-Hong}},\ }\bibfield
  {title} {\bibinfo {title} {Ashkin-teller formalism for elastic response of
  dna molecule to external force and torque},\ }\href
  {https://doi.org/10.1088/0253-6102/49/2/57} {\bibfield  {journal} {\bibinfo
  {journal} {Communications in Theoretical Physics}\ }\textbf {\bibinfo
  {volume} {49}},\ \bibinfo {pages} {525} (\bibinfo {year} {2008})}\BibitemShut
  {NoStop}%
\bibitem [{\citenamefont {Ashkin}\ and\ \citenamefont
  {Teller}(1943)}]{1943_ashkinteller}%
  \BibitemOpen
  \bibfield  {author} {\bibinfo {author} {\bibfnamefont {J.}~\bibnamefont
  {Ashkin}}\ and\ \bibinfo {author} {\bibfnamefont {E.}~\bibnamefont
  {Teller}},\ }\bibfield  {title} {\bibinfo {title} {Statistics of
  two-dimensional lattices with four components},\ }\href
  {https://doi.org/10.1103/PhysRev.64.178} {\bibfield  {journal} {\bibinfo
  {journal} {Phys. Rev.}\ }\textbf {\bibinfo {volume} {64}},\ \bibinfo {pages}
  {178} (\bibinfo {year} {1943})}\BibitemShut {NoStop}%
\bibitem [{\citenamefont {Carlon}\ \emph {et~al.}(2001)\citenamefont {Carlon},
  \citenamefont {Lajk\'o},\ and\ \citenamefont {Igl\'oi}}]{2001_ashkinteller}%
  \BibitemOpen
  \bibfield  {author} {\bibinfo {author} {\bibfnamefont {E.}~\bibnamefont
  {Carlon}}, \bibinfo {author} {\bibfnamefont {P.}~\bibnamefont {Lajk\'o}},\
  and\ \bibinfo {author} {\bibfnamefont {F.}~\bibnamefont {Igl\'oi}},\
  }\bibfield  {title} {\bibinfo {title} {Disorder induced cross-over effects at
  quantum critical points},\ }\href
  {https://doi.org/10.1103/PhysRevLett.87.277201} {\bibfield  {journal}
  {\bibinfo  {journal} {Phys. Rev. Lett.}\ }\textbf {\bibinfo {volume} {87}},\
  \bibinfo {pages} {277201} (\bibinfo {year} {2001})}\BibitemShut {NoStop}%
\bibitem [{\citenamefont {Goswami}\ \emph {et~al.}(2008)\citenamefont
  {Goswami}, \citenamefont {Schwab},\ and\ \citenamefont
  {Chakravarty}}]{2008_ashkinteller}%
  \BibitemOpen
  \bibfield  {author} {\bibinfo {author} {\bibfnamefont {P.}~\bibnamefont
  {Goswami}}, \bibinfo {author} {\bibfnamefont {D.}~\bibnamefont {Schwab}},\
  and\ \bibinfo {author} {\bibfnamefont {S.}~\bibnamefont {Chakravarty}},\
  }\bibfield  {title} {\bibinfo {title} {Rounding by disorder of first-order
  quantum phase transitions: Emergence of quantum critical points},\ }\href
  {https://doi.org/10.1103/PhysRevLett.100.015703} {\bibfield  {journal}
  {\bibinfo  {journal} {Phys. Rev. Lett.}\ }\textbf {\bibinfo {volume} {100}},\
  \bibinfo {pages} {015703} (\bibinfo {year} {2008})}\BibitemShut {NoStop}%
\bibitem [{\citenamefont {Hrahsheh}\ \emph {et~al.}(2014)\citenamefont
  {Hrahsheh}, \citenamefont {Hoyos}, \citenamefont {Narayanan},\ and\
  \citenamefont {Vojta}}]{2014_ashkinteller}%
  \BibitemOpen
  \bibfield  {author} {\bibinfo {author} {\bibfnamefont {F.}~\bibnamefont
  {Hrahsheh}}, \bibinfo {author} {\bibfnamefont {J.~A.}\ \bibnamefont {Hoyos}},
  \bibinfo {author} {\bibfnamefont {R.}~\bibnamefont {Narayanan}},\ and\
  \bibinfo {author} {\bibfnamefont {T.}~\bibnamefont {Vojta}},\ }\bibfield
  {title} {\bibinfo {title} {Strong-randomness infinite-coupling phase in a
  random quantum spin chain},\ }\href
  {https://doi.org/10.1103/PhysRevB.89.014401} {\bibfield  {journal} {\bibinfo
  {journal} {Phys. Rev. B}\ }\textbf {\bibinfo {volume} {89}},\ \bibinfo
  {pages} {014401} (\bibinfo {year} {2014})}\BibitemShut {NoStop}%
\bibitem [{\citenamefont {Chatelain}\ and\ \citenamefont
  {Voliotis}(2016)}]{chatelain-voliotis-epjb16}%
  \BibitemOpen
  \bibfield  {author} {\bibinfo {author} {\bibfnamefont {C.}~\bibnamefont
  {Chatelain}}\ and\ \bibinfo {author} {\bibfnamefont {D.}~\bibnamefont
  {Voliotis}},\ }\bibfield  {title} {\bibinfo {title} {Numerical evidence of
  the double-griffiths phase of the random quantum ashkin-teller chain},\
  }\href {https://doi.org/10.1140/epjb/e2015-60593-3} {\bibfield  {journal}
  {\bibinfo  {journal} {Eur. Phys. J. B}\ }\textbf {\bibinfo {volume} {89}},\
  \bibinfo {pages} {18} (\bibinfo {year} {2016})}\BibitemShut {NoStop}%
\bibitem [{\citenamefont {Caldeira}\ and\ \citenamefont
  {Leggett}(1981)}]{caldeiraleggett}%
  \BibitemOpen
  \bibfield  {author} {\bibinfo {author} {\bibfnamefont {A.~O.}\ \bibnamefont
  {Caldeira}}\ and\ \bibinfo {author} {\bibfnamefont {A.~J.}\ \bibnamefont
  {Leggett}},\ }\bibfield  {title} {\bibinfo {title} {Influence of dissipation
  on quantum tunneling in macroscopic systems},\ }\href
  {https://doi.org/10.1103/PhysRevLett.46.211} {\bibfield  {journal} {\bibinfo
  {journal} {Phys. Rev. Lett.}\ }\textbf {\bibinfo {volume} {46}},\ \bibinfo
  {pages} {211} (\bibinfo {year} {1981})}\BibitemShut {NoStop}%
\bibitem [{\citenamefont {Leggett}\ \emph {et~al.}(1987)\citenamefont
  {Leggett}, \citenamefont {Chakravarty}, \citenamefont {Dorsey}, \citenamefont
  {Fisher}, \citenamefont {Garg},\ and\ \citenamefont
  {Zwerger}}]{leggett_spinboson}%
  \BibitemOpen
  \bibfield  {author} {\bibinfo {author} {\bibfnamefont {A.~J.}\ \bibnamefont
  {Leggett}}, \bibinfo {author} {\bibfnamefont {S.}~\bibnamefont
  {Chakravarty}}, \bibinfo {author} {\bibfnamefont {A.~T.}\ \bibnamefont
  {Dorsey}}, \bibinfo {author} {\bibfnamefont {M.~P.~A.}\ \bibnamefont
  {Fisher}}, \bibinfo {author} {\bibfnamefont {A.}~\bibnamefont {Garg}},\ and\
  \bibinfo {author} {\bibfnamefont {W.}~\bibnamefont {Zwerger}},\ }\bibfield
  {title} {\bibinfo {title} {Dynamics of the dissipative two-state system},\
  }\href {https://doi.org/10.1103/RevModPhys.59.1} {\bibfield  {journal}
  {\bibinfo  {journal} {Rev. Mod. Phys.}\ }\textbf {\bibinfo {volume} {59}},\
  \bibinfo {pages} {1} (\bibinfo {year} {1987})}\BibitemShut {NoStop}%
\end{thebibliography}%

\end{document}